\documentclass[prd,12pt,aps,superscriptaddress,preprintnumbers,showpacs,tightenlines,floatfix,nofootinbib,amssymb]{revtex4}
\usepackage{epsfig}
\newcommand{\avr}{\langle L \rangle}
\newcommand{\avrtwo}{\langle L^2 \rangle}
\newcommand{\eff}{{\mathrm{eff}}}
\newcommand{\Abel}{{\mathrm{Abel}}}
\newcommand{\mon}{{\mathrm{mon}}}
\def\absc{(1-\vert c(s,\mu)\vert^2)^{1/2}}
\newcommand{\beq}{\begin{equation}}
\newcommand{\eeq}{\end{equation}}
\newcommand{\beqn}{\begin{eqnarray}}
\newcommand{\eeqn}{\end{eqnarray}}
\newcommand{\bea}[1]{\beq\begin{array}{#1}}
\newcommand{\eea}{\end{array}\eeq}
\newcommand{\eq}[1]{(\ref{#1})}

\newcommand{\cD}{{\cal D}}

\newcommand{\Kanazawa}{\affiliation{Institute for Theoretical Physics,
Kanazawa University, Kanazawa 920-1192, Japan}}

\newcommand{\ITEP}{\affiliation{Institute of Theoretical and
Experimental Physics, B.Cheremushkinskaya 25, Moscow, 117259, Russia}}

\begin{document}

\title{Numerical determination of monopole entropy \newline in pure SU(2) QCD}

\author{M.N.~Chernodub}\Kanazawa\ITEP
\author{Katsuya~Ishiguro}\Kanazawa
\author{Katsuya~Kobayashi}\Kanazawa
\author{Tsuneo~Suzuki}\Kanazawa

\preprint{KANAZAWA/2003-12}
\preprint{ITEP-LAT-2003-08}

\begin{abstract}
We study numerically the length distributions of the infrared monopole clusters in pure SU(2) QCD.
These distributions are Gaussian for all studied blocking steps of monopoles, lattice volumes and
lattice coupling constant. We also investigate the monopole action for the infrared monopole clusters.
The knowledge of both the length distribution and the monopole action allows us to determine the effective
entropy of the monopole currents. The entropy is a descending function of blocking scale,
indicating that the effective degrees of freedom of the extended monopoles are getting smaller
as the blocking scale increases.
\end{abstract}


\date{}

\maketitle

\section{Introduction}
\label{one}

The dual superconductor picture~\cite{DualSuperconductor} of the QCD vacuum
is one of the most promising approaches to the problem of color confinement.
This picture is based on the existence of Abelian monopoles in
the vacuum of QCD. The monopoles are identified with the help
of the so-called Abelian projection method~\cite{AbelianProjections}, which
is based on the partial gauge fixing of the SU(N) gauge symmetry up to an
Abelian subgroup. The monopoles naturally appear in the Abelian projection
due to compactness of the residual Abelian group.

There are various numerical indications that the monopoles are responsible
for the confinement of quarks (for a review, see Ref.~\cite{Reviews}). One
of the most important observations is the monopole condensation in the
low temperature (confinement) phase~\cite{shiba:condensation,MonopoleCondensation}.
According to the dual superconductor mechanism the monopole condensation give
rise to the formation of the chromoelectric string
which confines the fundamental color sources. This expectation is confirmed by the fact
that the non--zero tension of the chromoelectric string is dominated by the
Abelian monopole contributions~\cite{AbelianDominance,shiba:string,koma:string}.

In the numerical simulations one observes that the trajectories of the Abelian
monopoles form clusters, which can be divided by two ensembles:
finite-sized clusters and one large percolating cluster~\cite{ivanenko,ref:kitahara,ref:hart}.
The percolating cluster (or, infrared cluster) occupies the whole lattice while the
sizes of the other clusters have an ultraviolet nature. The existence of the
IR cluster is related to the monopole condensation~\cite{ivanenko}. This is
understandable on an intuitive level, since generally the condensation is a microscopic
effect reflecting itself in a zero--momentum component of the condensed field. The importance
of the IR cluster for the confinement of quarks was also stressed in numerical
calculations~\cite{ref:kitahara}: the tension of the confining string gets a dominant contribution
from the monopoles belonging to the IR cluster, while the contribution of the UV clusters to the string tension
is negligible. In the deconfinement phase the IR cluster disappears~\cite{ivanenko,ref:kitahara}, as
expected.

In this paper we mostly concentrate on the numerical investigation of the properties of the
infrared monopole cluster. The length distributions and other properties of the UV and IR clusters were
studied previously in Refs.~\cite{ref:zakharov:clusters,ref:hart,ref:kitahara,ref:boyko,zakharov:recent}.
In this publication we are going to investigate thoroughly the properties of the length distributions of
the monopole clusters for various lattice volumes and sizes of the extended monopoles.

The plan of the paper is the following. In Section~\ref{sec:model} we describe the model
and provide the details of numerical simulations.
Section~\ref{sec:action} is devoted to the investigation of the Abelian monopole action obtained by the
inverse Monte-Carlo method. The distribution of the cluster length in the infrared clusters is
studied in Section~\ref{sec:length}. The knowledge of the monopole action and cluster distribution allows
us, for the first time, to calculate the entropy of the lattice monopoles of various sizes.
Our conclusions are presented in the last Section.

\section{Model and simulation details}
\label{sec:model}

We study the pure SU(2) gluodynamics with the lattice Wilson action,
$ S(U) = - \frac{\beta}{2} {\mathrm{Tr}} U_P$,
where $\beta$ is the coupling constant and $U_P$ is the SU(2) plaquette constructed
from the link fields. All our results are obtained in the Maximal Abelian (MA)
gauge~\cite{kronfeld} which is defined by the maximization of the lattice functional
\beqn
R = \sum_{s,\hat\mu}{\rm Tr}\Big(\sigma_3 \widetilde{U}(s,\mu)
\sigma_3 \widetilde{U}^{\dagger}(s,\mu)\Big)\,,
\label{R}
\eeqn
with respect to the gauge transformations
$U(s,\mu) \to \widetilde{U}(s,\mu)=\Omega(s)U(s,\mu)\Omega^\dagger(s+\hat\mu)$.
The local condition of maximization can be written in the continuum limit as the
differential equation $(\partial_{\mu}+igA_{\mu}^3)(A_{\mu}^1-iA_{\mu}^2)=0$.
Both this condition and the functional \eq{R} are invariant under
residual U(1) gauge transformations, $\Omega^{\mathrm{Abel}}(\omega)
= {\mathrm{diag} (e^{i \omega(s)},e^{- i \omega(s)})}$.

The next step is Abelian projection of non--Abelian link variables to the
Abelian ones after the gauge fixing is done. An Abelian gauge field is
extracted from the $SU(2)$ link variables as follows:
\beqn
 \widetilde{U}(s,\mu) = \left( \begin{array}{cc}
         \absc        & -c^*(s,\mu) \\
                  c(s,\mu) &  \absc
\end{array} \right)
\left( \begin{array}{cc}
u(s,\mu) & 0 \\
0 & u^*(s,\mu)
\end{array} \right),
\label{eq:field:decomposition}
\eeqn
where $u(s,\mu)= \exp(i\theta(s,\mu))$ represents the Abelian link field
and $c(s,\mu)$ corresponds to charged matter fields.

The Abelian field strength $\theta_{\mu\nu}(s)\in(-4\pi,4\pi)$ is defined
on the lattice plaquettes by a link angle $\theta(s,\mu)\in[-\pi,\pi)$
as $\theta_{\mu\nu}(s)=\theta(s,\mu)+
\theta(s+\hat\mu,\nu)-\theta(s+\hat\nu,\mu)-\theta(s,\nu)$.
The field strength $\theta_{\mu\nu}(s)$ can be decomposed into two parts,
\beqn
\theta_{\mu\nu}(s)= \bar{\theta}_{\mu\nu}(s) +2\pi m_{\mu\nu}(s)\,,
\label{eq:field:separation}
\eeqn
where $\bar{\theta}_{\mu\nu}(s)\in [-\pi,\pi)$ is interpreted as
the electromagnetic flux through the plaquette
and $m_{\mu\nu}(s)$ can be regarded as a number of the Dirac
strings piercing the plaquette.

The elementary monopole currents is conventionally constructed using the
DeGrand-Toussaint\cite{degrand} definition:
\beqn
k_{\mu}(s) & = & \frac{1}{2}\epsilon_{\mu\nu\rho\sigma}
\partial_{\nu}m_{\rho\sigma}(s+\hat{\mu}),
\label{eq:monopole:definition}
\eeqn
where $\partial$ is the forward lattice derivative. The monopole current is defined
on a link of the dual lattice and takes values $0, \pm 1, \pm 2$. Moreover the
monopole current satisfies the conservation law automatically,
\beqn
\partial'_{\mu}k_{\mu}(s)=0\,,
\eeqn
where $\partial'$ is the backward derivative on the dual lattice.

Besides the elementary monopoles one can also define the so called
extended monopoles~\cite{ivanenko}. In this paper we use the type-2
construction according to the classification of the extended monopoles
adopted in Ref.~\cite{ivanenko}. The $n^3$ extended monopole is defined
on a sublattice with the lattice spacing $b=na$, where $a$ is
the spacing of the original lattice. Thus the construction of the
extended monopoles corresponds to a block spin transformation of the monopole
currents with the scale factor $n$,
\beqn
k_{\mu}^{(n)}(s) = \sum_{i,j,l=0}^{n-1}k_{\mu}(n s+(n-1)\hat{\mu}+i\hat{\nu}
     +j\hat{\rho}+l\hat{\sigma})\,.
\label{eq:blocking}
\eeqn

The Abelian dominance and the monopole dominance in the infrared region
of QCD implies that at least important infrared observables
(such as the fundamental string tension) can be calculated using the Abelian
fields or the monopole degrees of freedom only.

In what follows we discuss an
effective model of the monopole currents corresponding
to pure SU(2) QCD. Formally, we get this effective model through the gauge fixing procedure
applied to the original model. Then we integrate out of all degrees of freedom but
the monopole ones. An effective Abelian action is related to
the original non-Abelian action $S[C, \theta]$ (matter, $C$, and
Abelian gauge, $\theta$, fields, Eq.~\eq{eq:field:decomposition}) as follows:
\beqn
Z = \int \cD u \Bigl[\int DC e^{-S[C, \theta]} \delta(X) \Delta_{FP}(U)\Bigr]
= \int \cD u \, e^{-S_{eff}[\theta]}\,.
\label{eq:Z1}
\eeqn
Here and below we omit irrelevant constant terms in front of the partition functions.
In Eq.~\eq{eq:Z1} the term $\delta(X)$ represents the gauge-fixing condition\footnote{As we have
discussed above, the MA gauge fixing condition is given by a maximization of the functional \eq{R}
and therefore the use of the local condition $X=0$, implied in Eq.~\eq{eq:Z1}, is a formal
simplified notation.} and $\Delta_{FP}(U)$ is the corresponding Faddeev-Popov determinant.
Next step is to relate the effective monopole action to the effective U(1) action:
\beqn
Z  &=& \Bigl( \prod_{s, \mu} \sum_{k_\mu(s)
    = -\infty}^{\infty} \Bigr)
    \int \cD \theta \delta \bigl( k_\mu(s) - k_\mu(s;\theta)) \bigr)
    e^{-S^{\Abel}_{\eff}[\theta]} \\
   &=& \Bigl( \prod_{s, \mu} \sum_{k_\mu(s) = -\infty}^{\infty} \Bigr)
        \Bigl( \prod_s \delta_{ \partial_{\mu}^{\prime} k_\mu (s), 0} \Bigr)
        e^{-S^{\mon}_{\eff}[k]}\,,
\eeqn
where $k_\mu(s;\theta)$ is the monopole current defined as a function of the Abelian
fields, $\theta$, via relations~(\ref{eq:field:separation})  and (\ref{eq:monopole:definition}).

Our simulation statistics is represented in Table~\ref{tbl:simulation:statistics}. The gauge
configurations were generated with the help of the standard Monte-Carlo algorithm.
\begin{table}
\begin{center}
 \begin{tabular}{|r|r|r|r|}             \hline
  \multicolumn{1}{|c|}{Lattice} &\multicolumn{1}{c}{$\beta$}
  &\multicolumn{1}{|c|}{Blocking}&\multicolumn{1}{|c|}{Configuration}\\
  \multicolumn{1}{|c|}{size} &\multicolumn{1}{c}{}
  &\multicolumn{1}{|c|}{factor}&\multicolumn{1}{|c|}{number}\\
  \hline
  6  & 2.1$\sim$2.4 & 1 & 3000 \\
  8  & 2.1$\sim$2.4 & 1 & 3000 \\
  10 & 2.1$\sim$2.4 & 1 & 3000 \\
  12 & 2.1$\sim$2.4 & 2 & 3000 \\
  14 & 2.1$\sim$2.4 & 1 & 3000 \\
  16 & 2.1$\sim$2.4 & 2 & 3000 \\
  24 & 2.1$\sim$2.4 & 2,3,4 & 3000 \\
  32(SA) & 2.1$\sim$2.6 & 2,3 & 950 \\
  48 & 2.1$\sim$2.6 & 2,3,4,6,8 & 2200 \\ \hline
 \end{tabular}
 \end{center}\caption{Simulation statistics.}
\label{tbl:simulation:statistics}
\end{table}
In most simulations we use the usual iterative algorithm to fix the MA gauge. However,
in order to check the Gribov copy dependence of the MA gauge fixing we also use
the so called simulated annealing (SA) algorithm with five Gribov copies. We refer a reader for a detailed
description of the SA method to Ref.~\cite{ref:bali}, where the advantages of the
SA method compared to the iterative algorithm are illustrated.

\section{Monopole action for various clusters}
\label{sec:action}

It is well known that in gluodynamics the monopole trajectories can be separated into
the infrared and ultraviolet monopole clusters. There is only one IR monopole cluster which
occupies all volume of the lattice, and a large number of shorter monopole trajectories
(UV clusters).
\begin{figure}[!thb]
\vskip 12mm
\begin{tabular}{cc}
\includegraphics[angle=-00,scale=0.3]{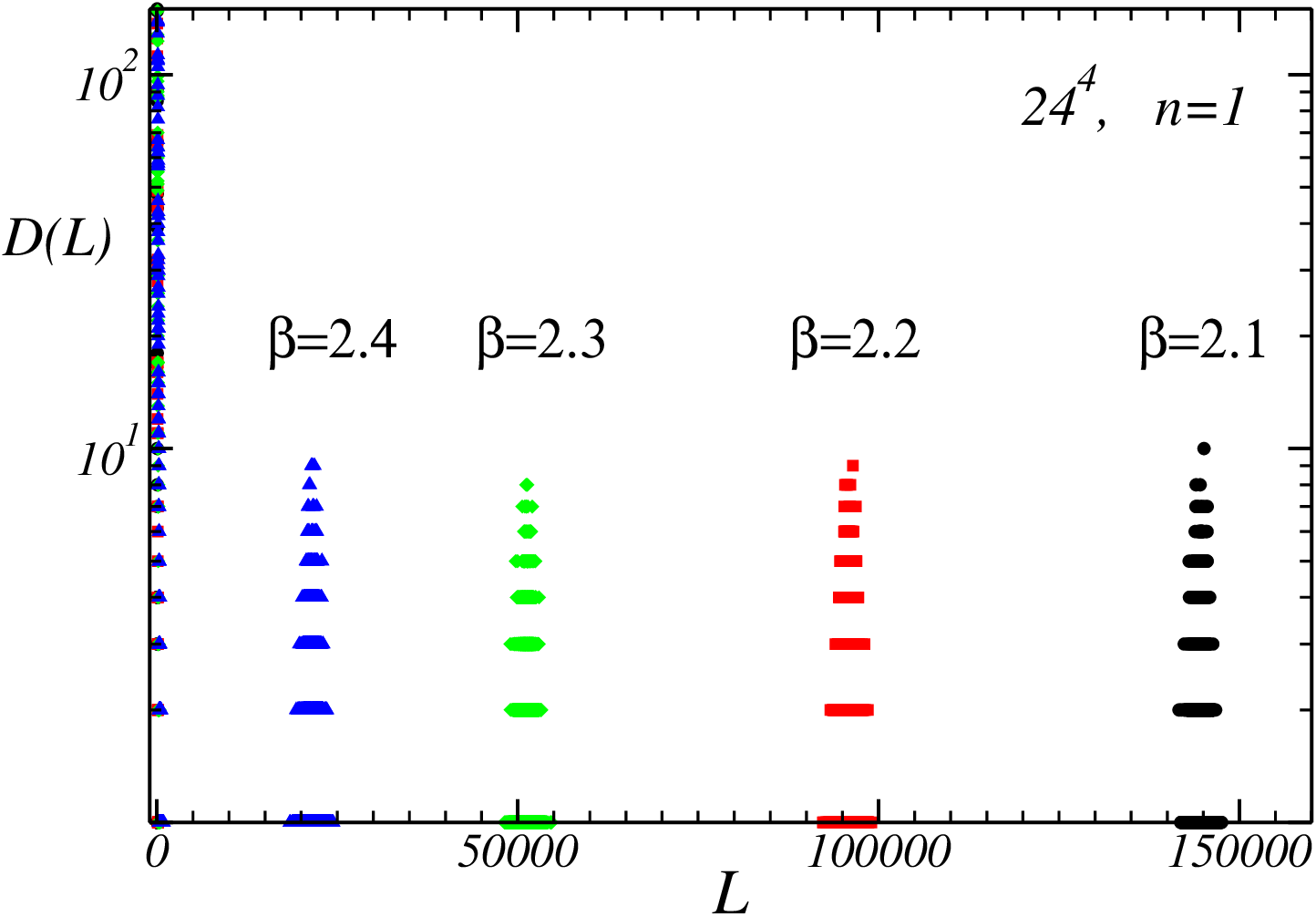}
&
\hskip 12mm\includegraphics[angle=-0,scale=0.3]{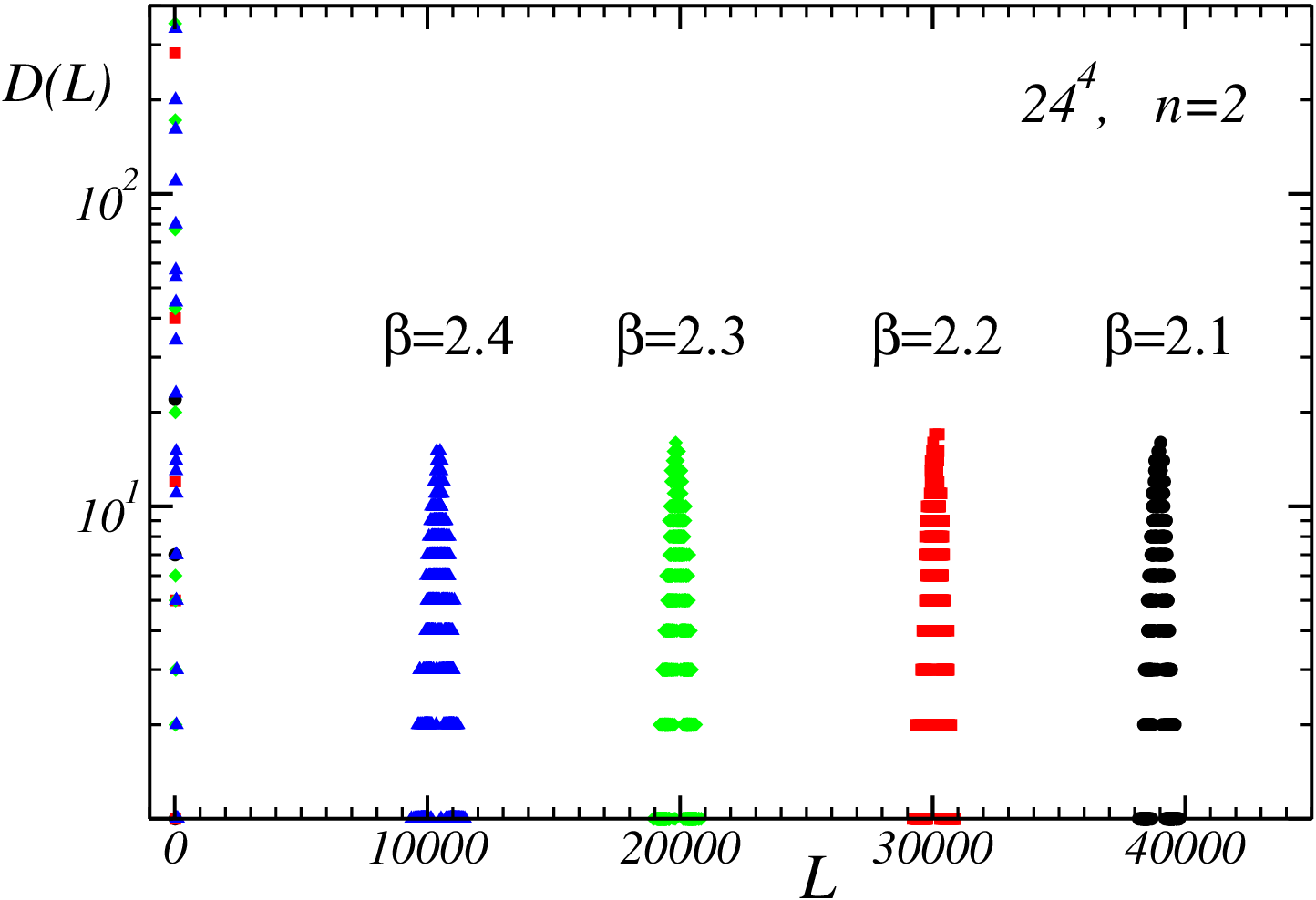}
\\
(a) & (b) \\
\end{tabular}
\caption{Distribution of the lengths of the monopole trajectories at various $\beta$ for
(a) elementary and (b) $n=2$ blocked monopoles.}
\label{fig:distribution:raw}
\end{figure}
In Figure~\ref{fig:distribution:raw} we show the typical length distributions, $D(L)$,
of the monopole trajectories in all clusters. We show the data for
elementary, $n=1$, and blocked, $n=2$, monopoles at various lattice coupling constants
$\beta$. One can see that for all considered values of the coupling $\beta$ the infrared cluster
and the ultraviolet clusters can be unambiguously separated due to a wide gap between them.
Moreover, the distributions of the elementary and blocked monopoles are qualitatively similar.

Note that at zero temperature
the gap between IR and UV clusters becomes smaller as the {\it physical} lattice size decreases.
This behaviour can be observed in Figure~\ref{fig:distribution:raw}.
At very small lattice size the gap between UV and IR clusters disappears and the IR and UV
clusters can not be resolved. This behaviour of the monopole clusters leads to the deconfining
transition ("crossover") which takes place in sufficiently small physical volumes.

The distribution of the ultraviolet clusters was studied
both numerically~\cite{ref:hart,zakharov:recent} and analytically~\cite{ref:zakharov:clusters,ref:zakharov}.
The distribution can be described by a power law $D_{UV} \propto L^{-\tau}$,
where the power $\tau$ is very close to 3, Ref.~\cite{ref:hart}. This behaviour
indicates that the monopoles in UV clusters show a random walk picture~\cite{ref:zakharov:clusters}.
In our simulations we are mainly concentrated on the IR monopole cluster because, as we have
already mentioned in the Introduction, the IR cluster is important for the confinement of quarks.
Below we study the monopole--related quantities using the largest monopole cluster only unless
stated otherwise.

In general, the monopole action, $S^{\mon}_{\eff}$, can be represented as a sum of
the $n$--point ($n \ge 2$) operators $S_i$, Ref.~\cite{shiba:condensation,chernodub}:
\beqn
S[k] = \sum_i f_i S_i [k]\,,
\label{eq:monopole:action}
\eeqn
where $f_i$ are coupling constants. In this paper we adopt only the two--point interactions
in the monopole action ($i.e.$ interactions of the form $S_i \sim k_{\mu}(s) k_{\mu'}(s')$).
Following Ref.~\cite{shiba:condensation} we derive the effective monopole action~\eq{eq:monopole:action}
from the configurations of monopole currents,
$\{ k_\mu (s) \}$ using an inverse Monte-Carlo method. The original monopole configurations
were generated by the usual Monte-Carlo algorithm of SU(2) gluodynamics.

\begin{figure}[thb]
\vskip 12mm
\begin{tabular}{cc}
\includegraphics[angle=0,scale=0.3]{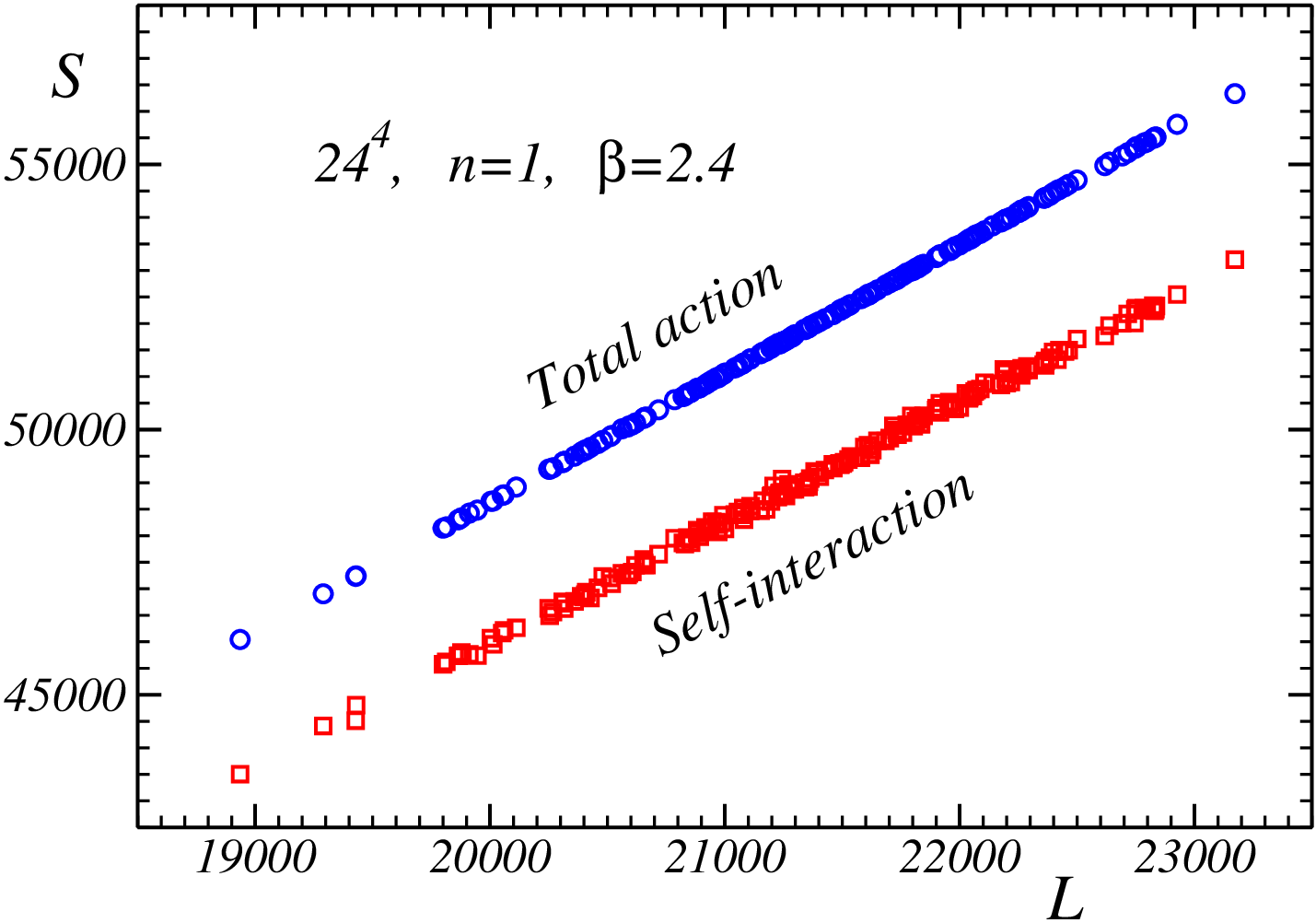} &
\hskip 12mm
\includegraphics[angle=0,scale=0.3]{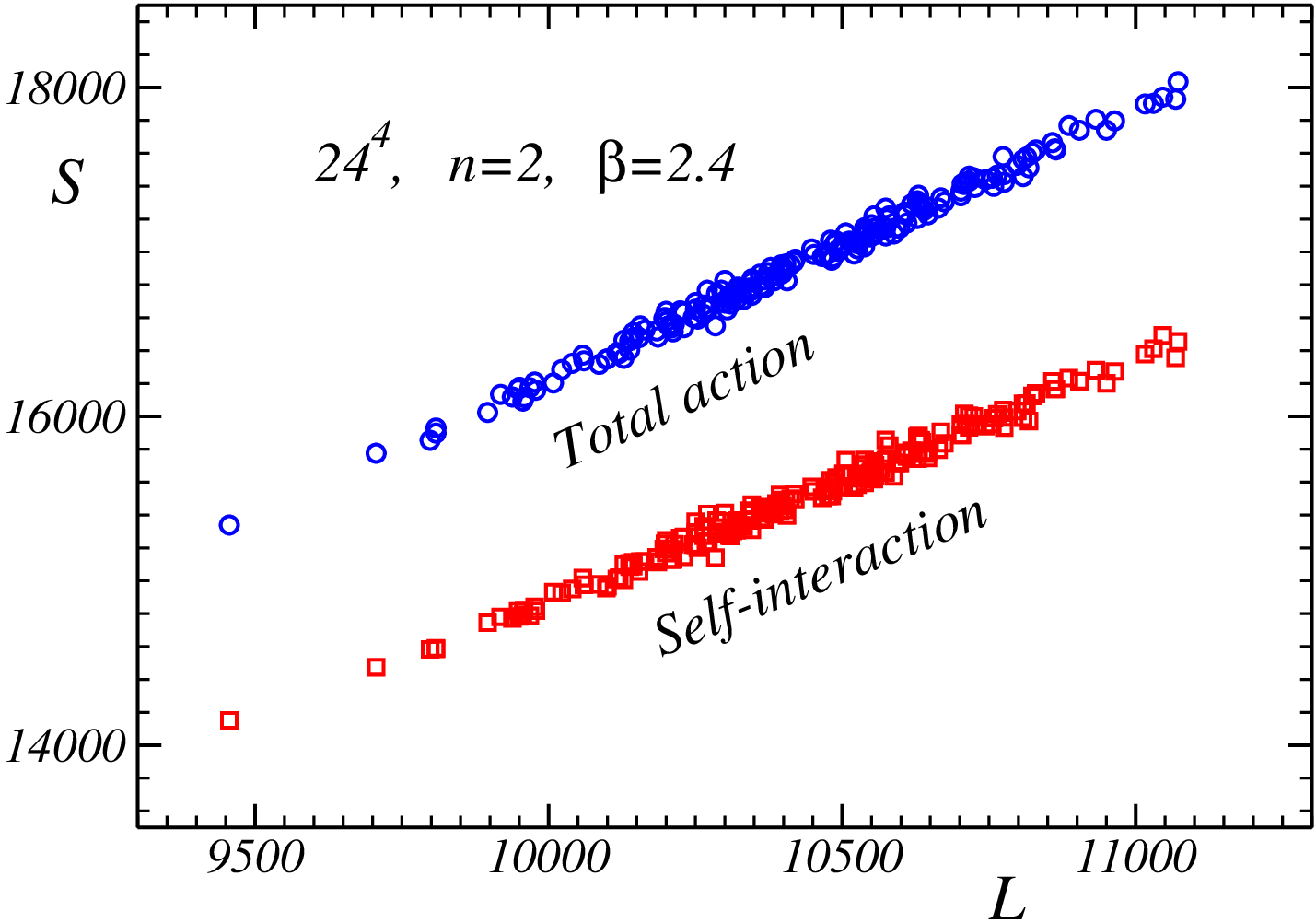}\\
(a) & (b) \\
\end{tabular}
\caption{The total monopole action and the contribution of the self--interaction term to the action
for (a) elementary and (b) $n=2$ extended monopoles $vs.$ length of the monopole trajectory
in the IR cluster as calculated on $24^4$ lattice at $\beta=2.4$.}
\label{fig:action:total:self}
\end{figure}

The dominant term in the monopole action~\eq{eq:monopole:action} corresponds to the most
local self-interaction of the monopole currents, $S_1[k] = \sum_{s,\mu} k^2_\mu(s)$. The
contributions to the action associated with other interactions are small compared to the
leading term. As an example we show the leading contribution and the full action associated
with the IR monopole cluster for $\beta=2.4$ and $n=1,2$ in Figure~\ref{fig:action:total:self}.
Moreover, one can find that both the monopole action and the self--coupling contribution to it
are proportional with a good accuracy to the length of the monopole loop.

\begin{figure}[!thb]
\vskip 12mm
\includegraphics[angle=0,scale=0.4]{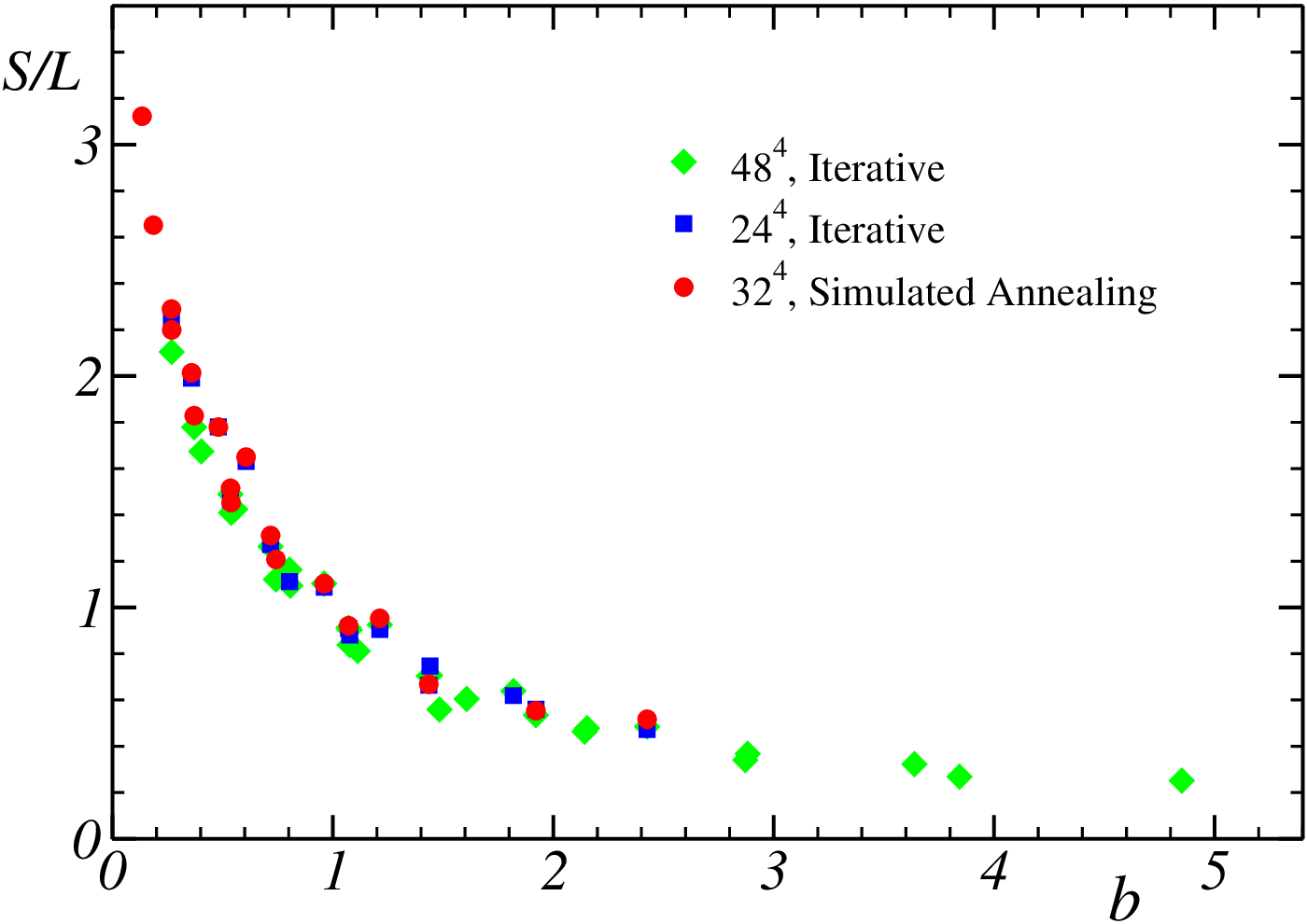}
\caption{The ratio $S\slash L$, in physical units, as the function of $b$ for various lattices, $N^4$,
and blocking steps, $n$.}
\label{fig:SL}
\end{figure}

In Figure~\ref{fig:SL} we plot the ratio $S[k]/L$ for various lattice volumes and blocking
sizes\footnote{In this figure and all other figures below we plot all dimensional quantities in units of the
string tension,~$\sigma$.}.
One can notice that the coupling constant depends on the product $b=a\cdot n$ and almost does not
depend on the variables $a$ and $n$ separately, in agreement with observations of
Ref.~\cite{shiba:condensation}. Below we will observe this type of scaling in many other
monopole quantities. Another observation is that
the monopoles obtained with the SA procedure have the same action as the monopoles
defined in the MA gauge which is fixed by the usual iterative algorithm.

It would also be interesting to compare the monopole action associated with
the IR cluster and the action associated with the whole monopole ensemble.
The simplest quantity to compare is
the $f_1$ self--coupling parameter which is a dominant coupling in the action.
\begin{figure}[thb]
\vskip 12mm
\includegraphics[angle=0,scale=0.4]{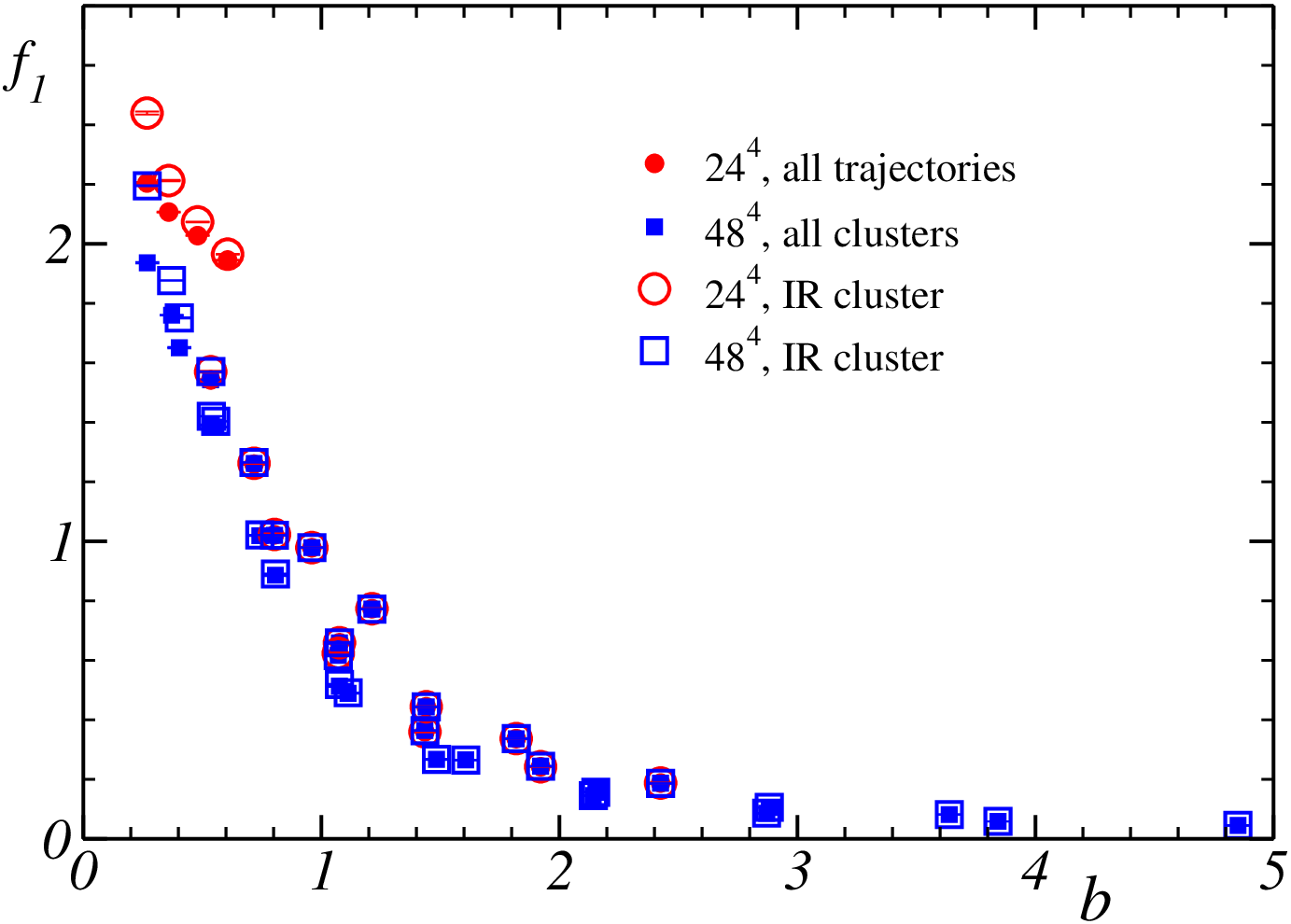}
\caption{The self--interaction coupling constant $f_1$ as the function of $b$
calculated for the largest monopole cluster and for the whole monopole ensemble on lattices
$24^4$ and $48^4$.}
\label{fig:f1}
\end{figure}
In Figure~\ref{fig:f1} we show $f_1$ for both ensembles. First, we easily notice that for
chosen lattices the coupling constant $f_1$ is independent of the lattice volume. Second,
we see that for large blocking scales $b$ the type of the ensemble (the IR cluster or the
whole ensemble) is not essential for determination of $f_1$. However, at small $b$ values,
$b \sqrt{\sigma} \lesssim 0.5$, the type of the lattice ensemble becomes important, since
in this region
\beqn
f^{\mathrm{IR}}_1 > f^{\mathrm{total}}_1\,,\quad \mbox{for}\,\,\,\, b \sqrt{\sigma} \lesssim 0.5\,.
\label{eq:f1:coupling}
\eeqn
The observed difference between the couplings can be affected by finite--size effects
since the leftmost points in our data correspond to elementary (of size $a$) monopoles.
Moreover, in our studies we have included only the two--point interactions in the monopole
action~\eq{eq:monopole:action}, while the two--point action becomes unreliable at too small
values of $b$, and one has to include higher--point interactions.

Despite a possible influence of the lattice artifacts, the observation~\eq{eq:f1:coupling}
may have a physical meaning related to the simple fact that the larger coupling $f_1$ the
smaller density of the monopoles is. Thus Eq.~\eq{eq:f1:coupling} is in agreement with the
numerical fact at large lattice coupling $\beta$ ($i.e.$, at small lattice spacing $a$)
the density of the monopoles in the largest cluster is
noticeably smaller than the total monopole density~\cite{ref:bornyakov}. Figure~\ref{fig:f1}
is also in a qualitative agreement with the fact~\cite{ref:anatomy} that
the excess of the Abelian action around elementary monopoles in all clusters and
in the IR cluster almost coincide with each other. However, the larger the physical size of
the monopole cube the better the agreement between the all--cluster and IR--cluster actions is, in
accordance with Figure~\ref{fig:f1}.

\section{Monopole length distribution for IR cluster}
\label{sec:length}

As we have mentioned, the length distribution of the monopole trajectories in the ultraviolet clusters
was found to obey the power--law. It would also be interesting to study the length distribution
for the infrared clusters following Ref.~\cite{ref:kitahara} .

Since the density of the elementary monopoles from infrared clusters is finite
(in terms of physical units) in the continuum limit~\cite{ref:bornyakov}, we may expect
that the density of the extended monopoles (with a fixed blocking scale $b$) is finite as well.
The finiteness of the density is consistent with the
observation that the monopole
length distribution is localized around a certain value of the monopole length, $L_{\mathrm{max}}$
(see Figure~\ref{fig:distribution:raw}). This value should be proportional to the
physical volume, $V$, of the system, $L_{max} \propto V$. Indeed, as one can qualitatively
judge from Figure~\ref{fig:distribution:raw}, the position of the peak of the IR length distribution
increases with increase of the physical volume of the system ($i.e.$, with decrease of
the lattice coupling $\beta$).

The length distribution function, $D(L)$, is proportional to the
weight with which the particular trajectory of the length $L$ contributes to the partition function.
On the other hand, the action of a monopole
trajectory is proportional to the length of the trajectory, $S \propto L$, as we have illustrated
in the previous Section. Thus the monopole action contributes in a form of an exponential factor,
$\propto e^{- f L}$, to the weight with which this trajectory appears in the
partition function. Here $f$ is a parameter which is close to the self-coupling $f_1$ according
to Figure~\ref{fig:f1}. The entropy of the monopole trajectory also contributes to the
monopole length distribution, which is proportional to $\mu^L$ (with $\mu$ being positive number) for
sufficiently large monopole lengths, $L$. Thus the distribution of the monopole trajectories in
infinite volume must be described by the function
\beqn
D^{IR}_{\inf}(L) \propto \mu^L \cdot e^{ - f \, L} = e^{\gamma L}\,, \quad \gamma = \ln \mu - f\,.
\label{eq:IR:distr:inf}
\eeqn
In this equation we neglect a power-law prefactor which is essential for the distribution of the
ultraviolet clusters\footnote{Below we work with the distribution of the pure exponential
form~\eq{eq:IR:distr:inf}. We also repeated our analysis with the prefactor $L^{-3}$ included.
We observed that the results with and without the power--law prefactor are the same
within the small error bars.}~\cite{ref:zakharov:clusters}.

The observed localization of the infrared cluster distribution imply that the
finite volume provide a certain cut which depends on the volume of the system. The simplest
distribution of this kind can be described by the function:
\beqn
D^{IR}(L) = \exp\{ - \alpha L^\eta + \gamma L\}\,,
\label{eq:IR:distr:eta}
\eeqn
where $\alpha$, $\gamma$ and $\eta$ are certain parameters. As we find below, the parameter
$\eta$ which characterizes the cut due to the volume effect, is close to 2 with
a big accuracy, $\eta \approx 2$. Moreover, as we mentioned, the parameter $\gamma$ is
characterizing the action and the entropy of the monopole currents and thus it should
depend only of the physical size of the blocked monopole, $\gamma = \gamma(b)$. As for the
parameter $\alpha$, it should also be dependent of the volume of the system,
$\alpha = \alpha(b,V)$. Thus we employ the following parameterization of the IR monopole
distribution at finite volume:
\beqn
D^{IR}(L) = \exp\{ - \alpha(b,V) L^2 + \gamma(b) L\}\,.
\label{eq:IR:distr:two}
\eeqn

The peak of the distribution \eq{eq:IR:distr:two},
\beqn
L_{max} = \frac{\gamma(b)}{2 \, \alpha(b,V)}\,,
\label{eq:Lmax}
\eeqn
is expected to be proportional to the volume of the system, $L_{max} \propto V$ to insure
the finiteness of the IR monopole density,
\beqn
\rho_{IR} = \frac{L_{max}}{V} = \frac{\gamma(b)}{2 \, \alpha(b,V)\, V}\,,
\label{eq:Lmax:2}
\eeqn
in the thermodynamic limit, $V \to \infty$. Thus from Eq.\eq{eq:Lmax:2} we conclude that
\beqn
\alpha(b,V) = A(b) \slash V\,,
\label{eq:alpha:b:V}
\eeqn
where the function $A(b)$ depends only on the size of the blocked monopole, $b$. Eq.\eq{eq:alpha:b:V}
implies that in the thermodynamic limit the parameter $\alpha$ vanishes and the finite--volume
distribution \eq{eq:IR:distr:eta} is reduced to Eq.\eq{eq:IR:distr:inf}, as expected.

\begin{figure}[thb]
\vskip 12mm
\begin{tabular}{cc}
\includegraphics[angle=-0,scale=0.3]{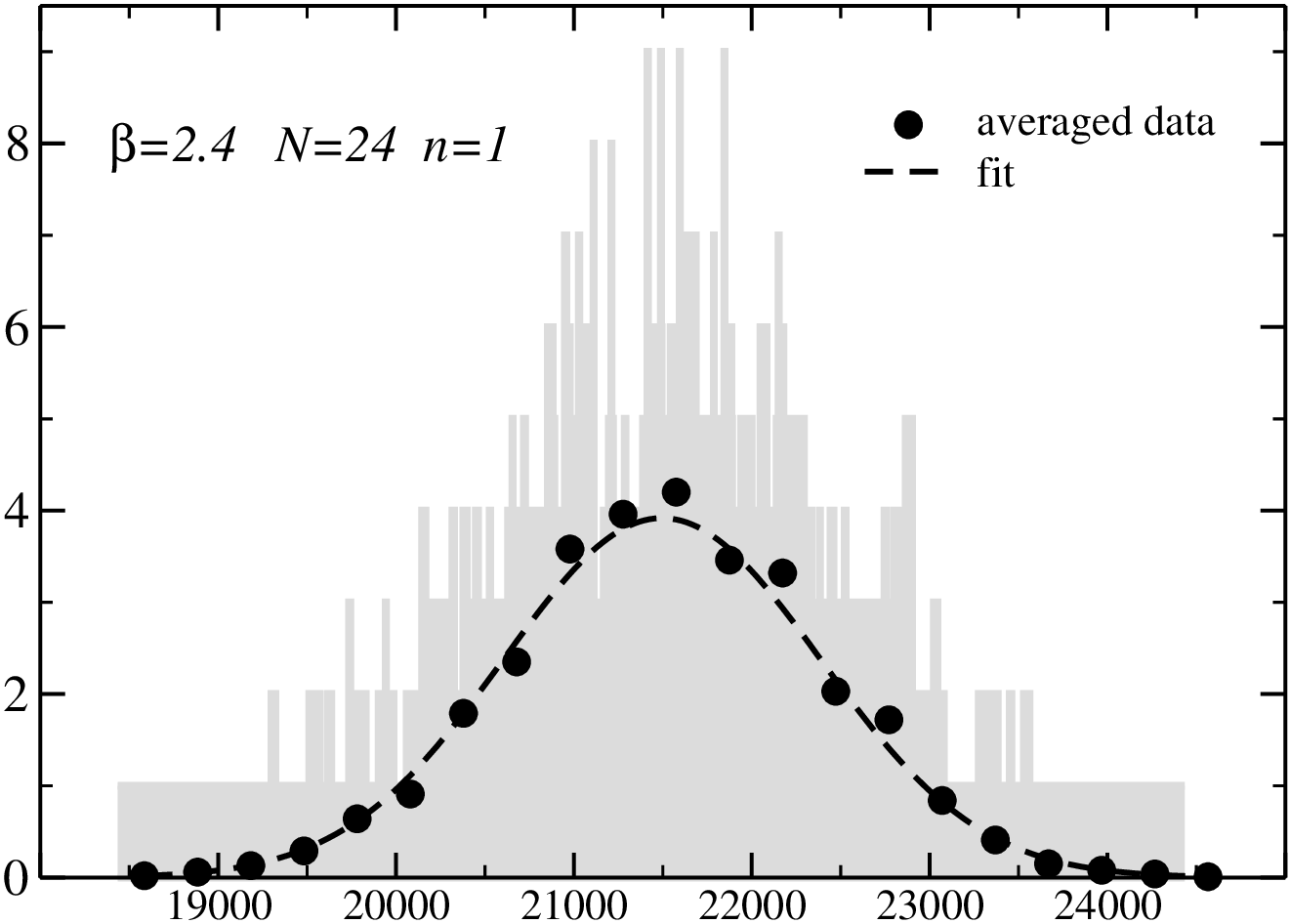} &
\hskip 12mm  \includegraphics[angle=-0,scale=0.3]{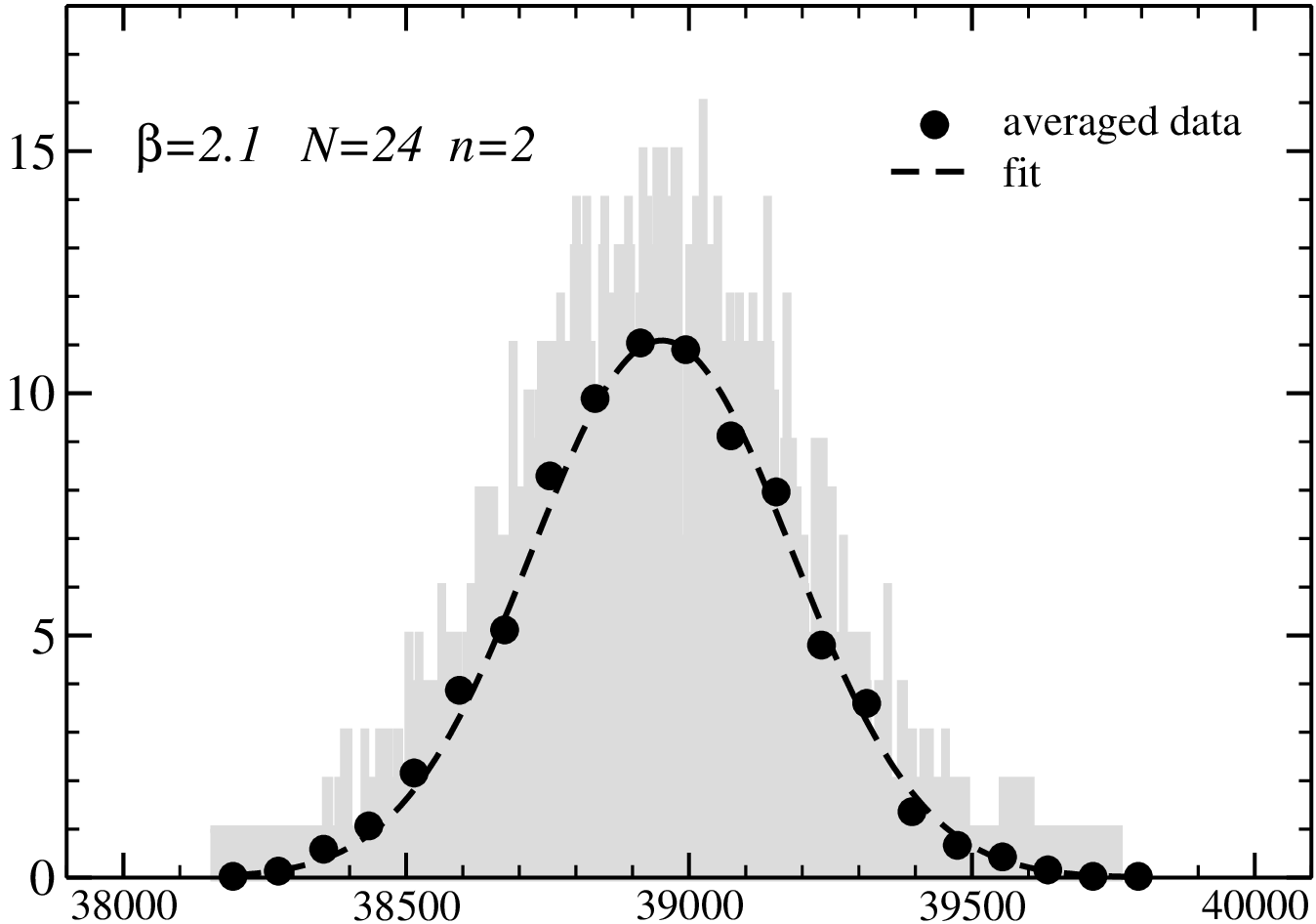}\\
(a) & (b) \\
\end{tabular}
\caption{The original histograms of the length distribution in the IR cluster
are shown by grey color. The averaged distributions are shown by circles, and the fits
by the function~\eq{eq:IR:distr:two} are represented by the dashed line.}
\label{fig:histogram}
\end{figure}
We show typical examples of the IR cluster distributions in Figure~\ref{fig:histogram} by grey color.
One can see that these histograms have an almost symmetric structure, but due to the lack of statistics
these histograms can not be fitted by the function~\eq{eq:IR:distr:two}. In order to show that
the distribution of the monopole follows Eq.\eq{eq:IR:distr:two} we smooth the data by increasing the
step of the histograms (which was equal to $2$) and then averaging the data inside the coarse steps.
We show the averaged (and suitably rescaled) histograms and their fits by the function~\eq{eq:IR:distr:two}
in the same Figure. One can see that the averaged histograms are very close to the Gaussian distribution.
Similar behaviour can also be observed for all IR monopole cluster distributions we have
studied in this paper.

In order to justify the chosen value of the parameter $\eta$ in Eq.~\eq{eq:IR:distr:two} we have also fitted the
averaged histogram data by Eq.~\eq{eq:IR:distr:eta} in which $\eta$ is treated as a fitting parameter.
The best fit (for $\beta=2.4$ and $n=1$ on $24^4$ lattice, as an example) gives us the result $\eta=2.05(15)$.
Fits of other histograms give us similar results. Thus we fix below $\eta=2$.

The histograms in Figure~\ref{fig:histogram} were obtained with a high simulation statistics
(3000 configurations according to Table~\ref{tbl:simulation:statistics}). In order to get
a perfect gaussian we would need
much more statistics which would consume a lot of CPU time. To avoid this we assume that the numerical
data for length distribution of the IR monopoles is described by Eq.\eq{eq:IR:distr:two}. This
assumption is justified by the analysis we have performed above. Then one can evaluate the central
values of the parameters $\alpha$ and $\gamma$ using the simple formulae (valid for a Gaussian
distribution):
\beqn
\alpha =  \frac{1}{2} \frac{1}{\avrtwo-\avr^2}\,,\quad
\gamma =  \frac{\avr}{\avrtwo-\avr^2}\,,
\label{eq:gauss:avr}
\eeqn
where the averaging $\langle \cdots \rangle$ is performed using weights from the histograms.

\begin{figure}[thb]
\vskip 12mm
\begin{tabular}{cc}
\includegraphics[angle=-0,scale=0.3]{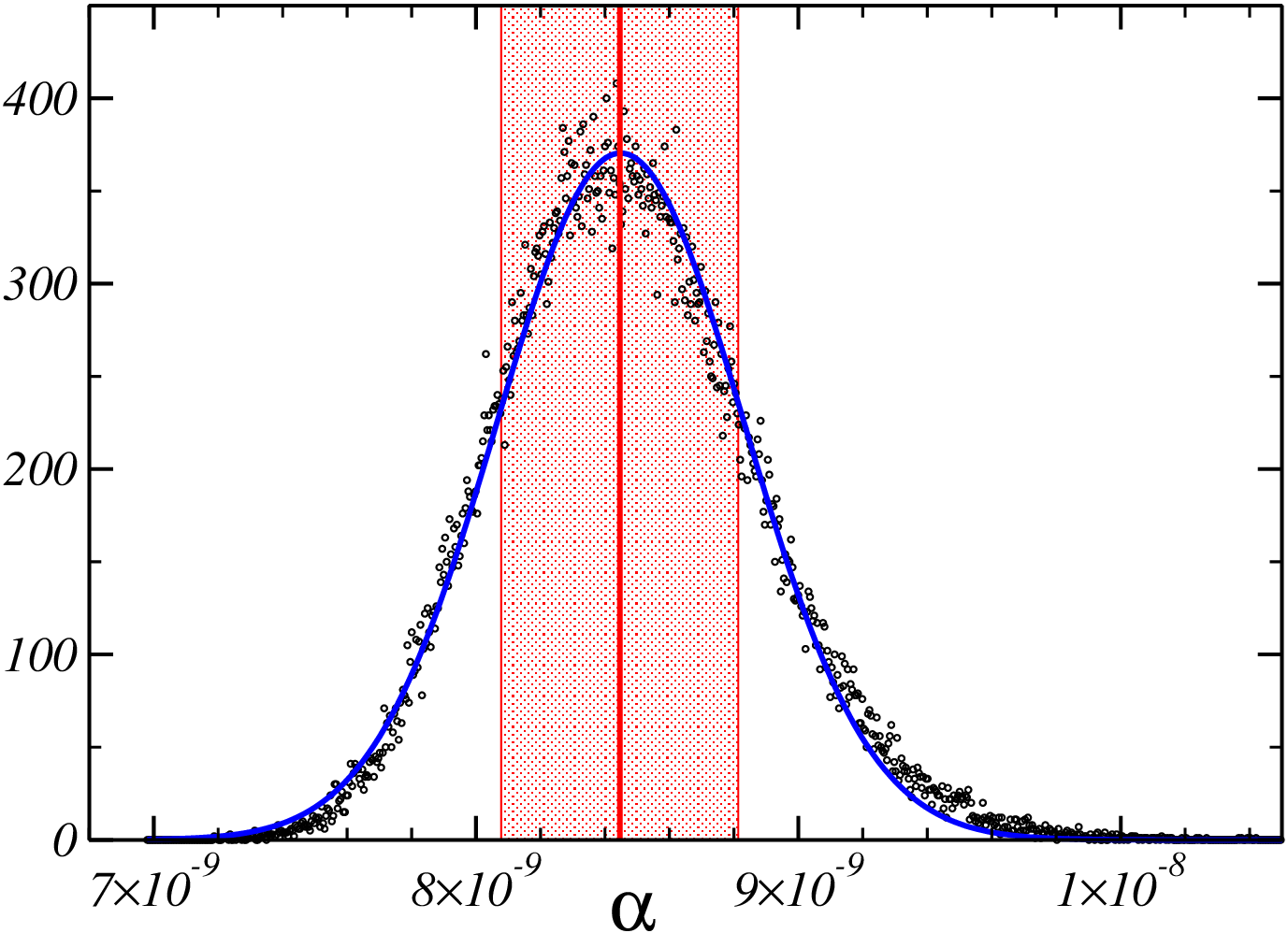} &
\hskip 12mm  \includegraphics[angle=-0,scale=0.3]{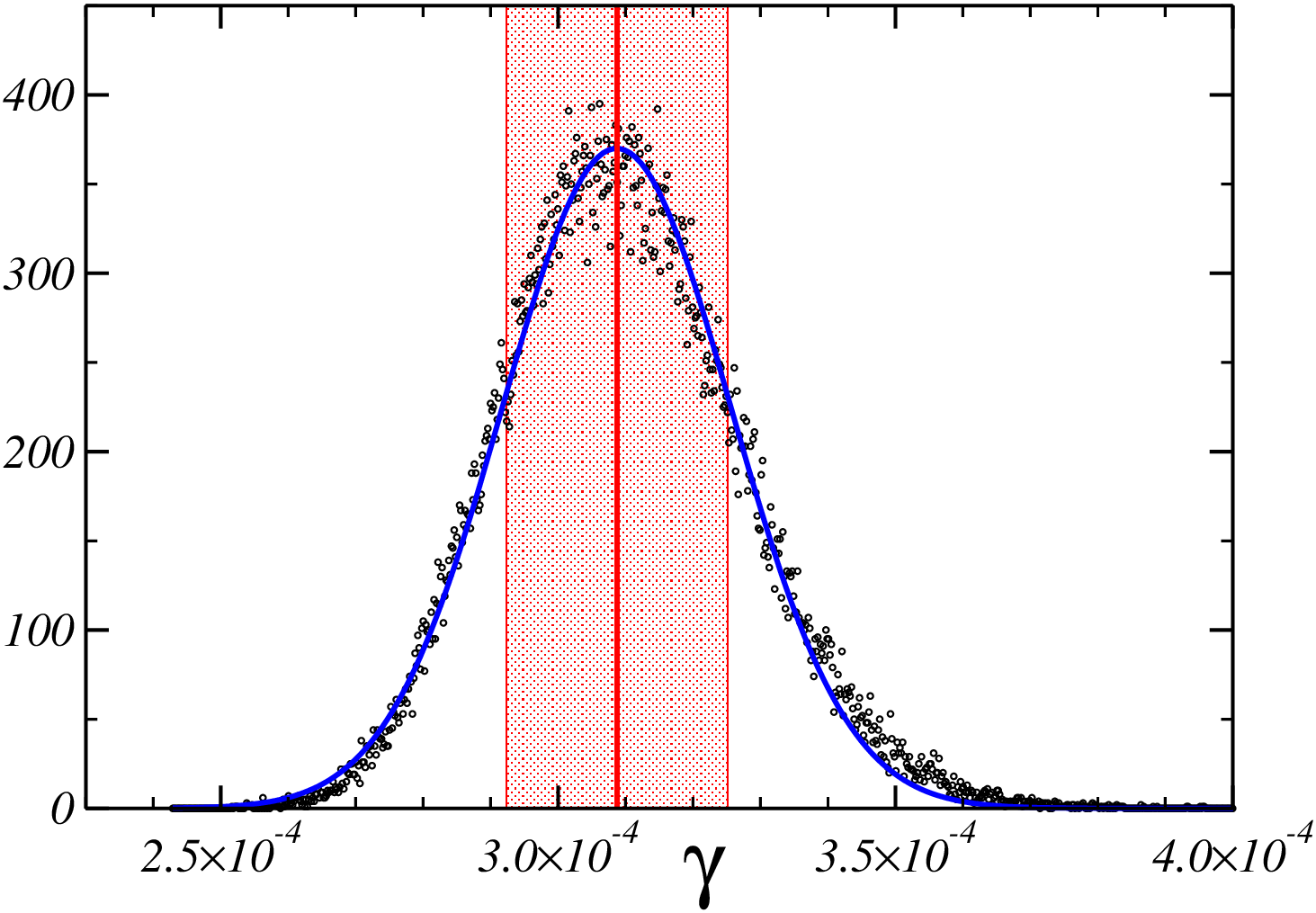}\\
(a) & (b) \\
\end{tabular}
\caption{The distribution of the parameter $\alpha$ and $\gamma$ for elementary monopoles
at $\beta=2.4$ on $24^4$ lattice. The fits by a gaussian function are shown by the solid lines
and the value of the errors are indicated by shadowed regions.}
\label{fig:alpha:beta:Gauss}
\end{figure}

To evaluate the errors for the parameters $\alpha$ and $\gamma$ we use the standard bootstrap method.
Namely, we make a resampling of the original data describing lengthes of the IR monopole clusters,
$L_{\max}$. We construct a resampled configuration by selecting $n_{conf}$ random values of $L_{\max}$
(note that a single value of $L_{\max}$ can be picked up multiple number of times),
where $n_{conf}$ is the total number of the monopole configurations. Then we evaluate the values of
$\alpha$ and $\gamma$ at each resampled configuration using Eq.~\eq{eq:gauss:avr}. The distribution of
these values is again the Gaussian with the width equal to the corresponding error. We plot examples of
the histograms for $\alpha$ and $\gamma$ values in Figures~\ref{fig:alpha:beta:Gauss}(a,b).

\begin{figure}[thb]
\vskip 12mm
\includegraphics[angle=-0,scale=0.4]{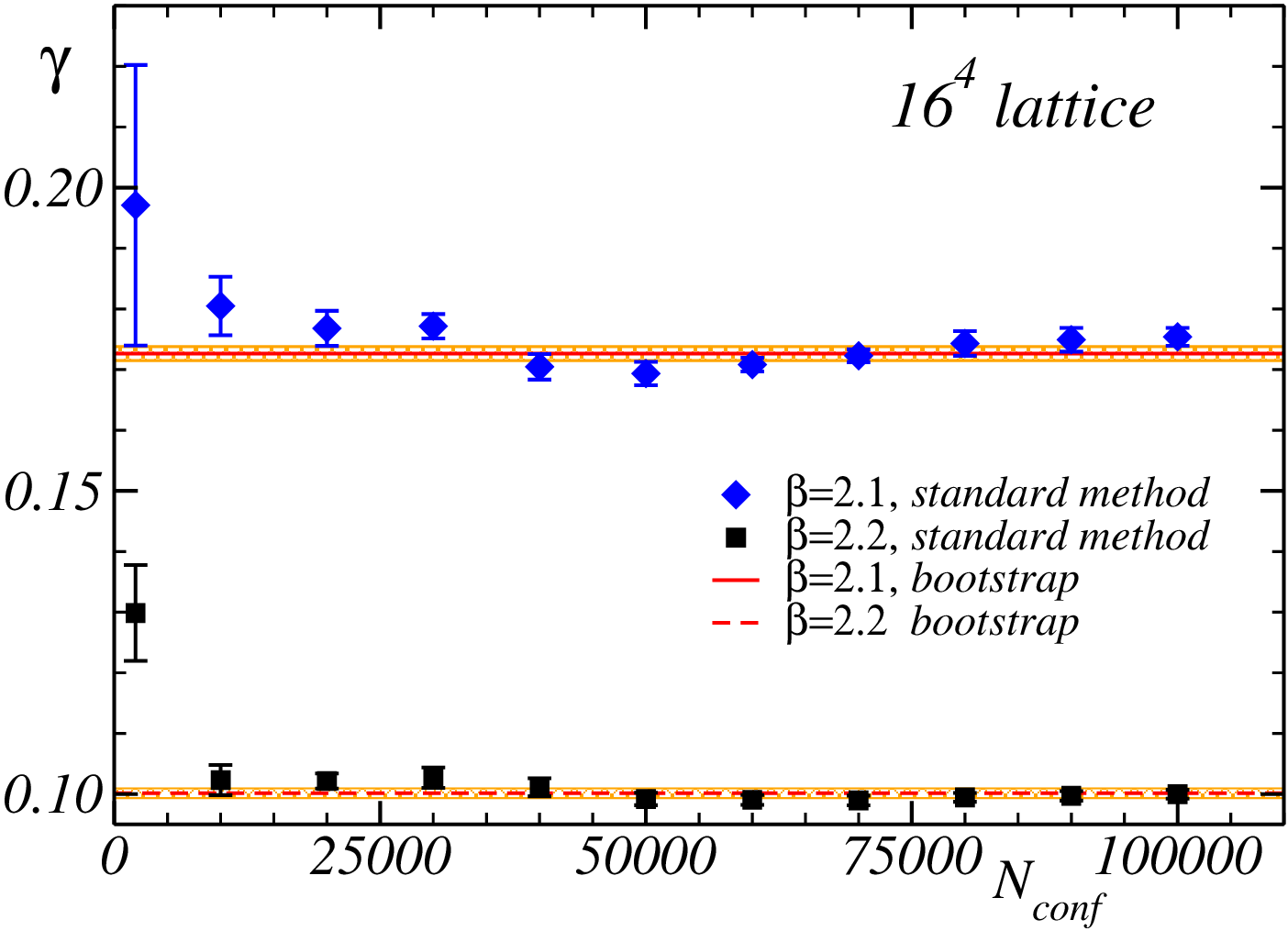}
\caption{Check of the bootstrap method on $1643$ lattice using the parameter $\gamma$ as an
example (the explanation is in the text).}
\label{fig:check:boot}
\end{figure}

We have checked the applicability of Eqs.~\eq{eq:gauss:avr} and the use of the bootstrap method
on a smaller, $16^4$, lattice. Namely, we have generated length distributions using
from low statistics ($2000$ configurations) to high statistics ($10^5$ configurations) ensembles.
We used the bootstrap method along
with Eqs.~\eq{eq:gauss:avr} to evaluate the coefficients $\alpha$ and $\beta$ for the distribution measured
with the lowest statistics. On the other hand, the high statistics distribution is a (almost perfect) Gaussian
and therefore we get the desired coefficients directly from the fit \eq{eq:IR:distr:two}.
The comparison of the coefficients shows that the central values as well as the estimated errors
for the low and for the high statistics ensembles coincide with each other within a few percents.
We illustrate our analysis in Figure~\ref{fig:check:boot} for $\beta=2.1$ and $\beta=2.2$
using the parameter $\gamma$ as an example. The values of $\gamma$ obtained with the standard method
are plotted $vs.$ number of configurations, $N_{conf}$, used in the analysis. The horizontal lines represent
the results coming from the bootstrap method applied to the low-statistic ensemble (the statistical errors
are indicated by shadowed regions). We conclude that
the bootstrap method allows to get reliable results using the distributions with low--statistics.

\begin{figure}[thb]
\vskip 12mm
\begin{tabular}{cc}
\includegraphics[angle=-0,scale=0.3]{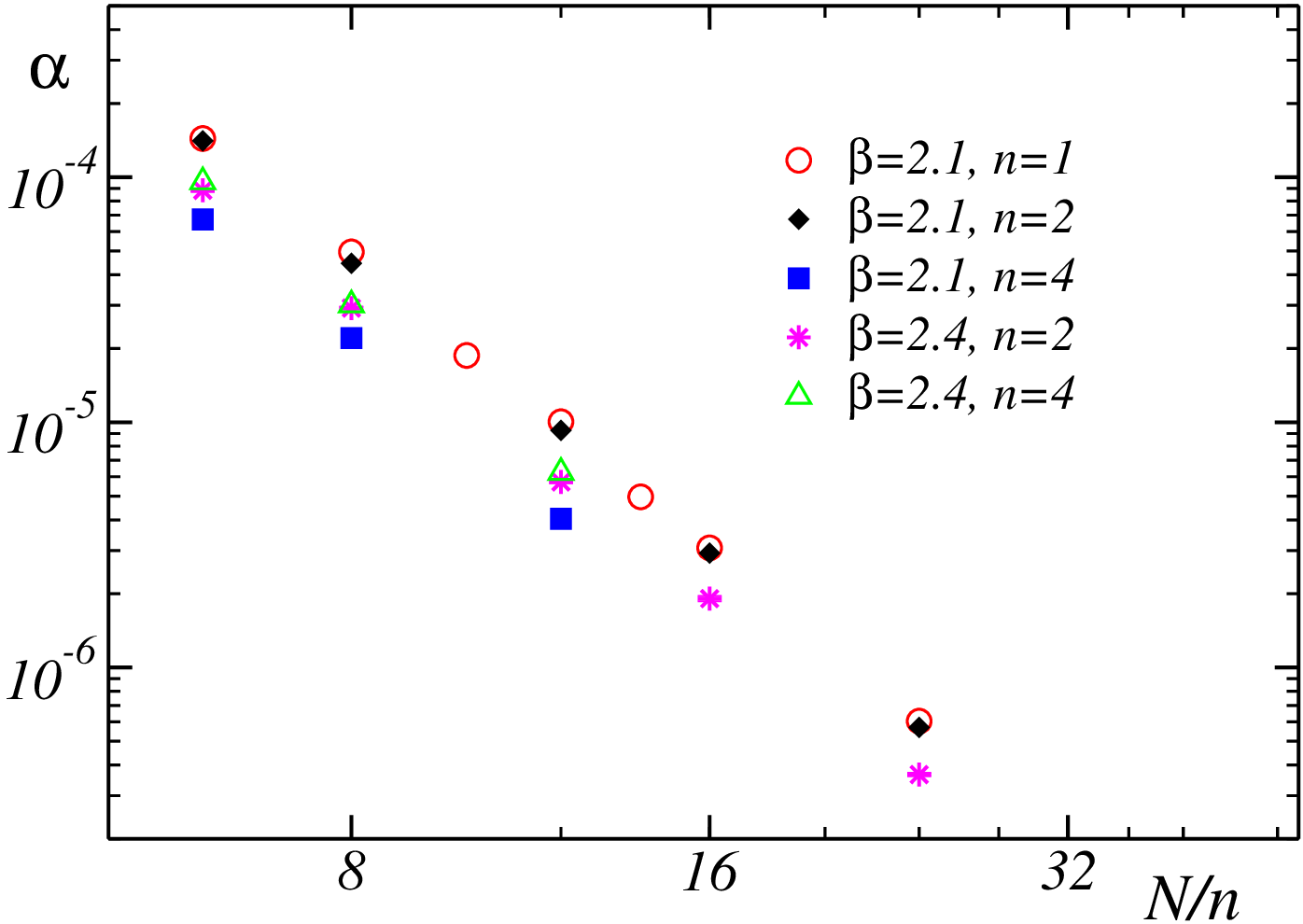} &
\hskip 12mm
\includegraphics[angle=-0,scale=0.3]{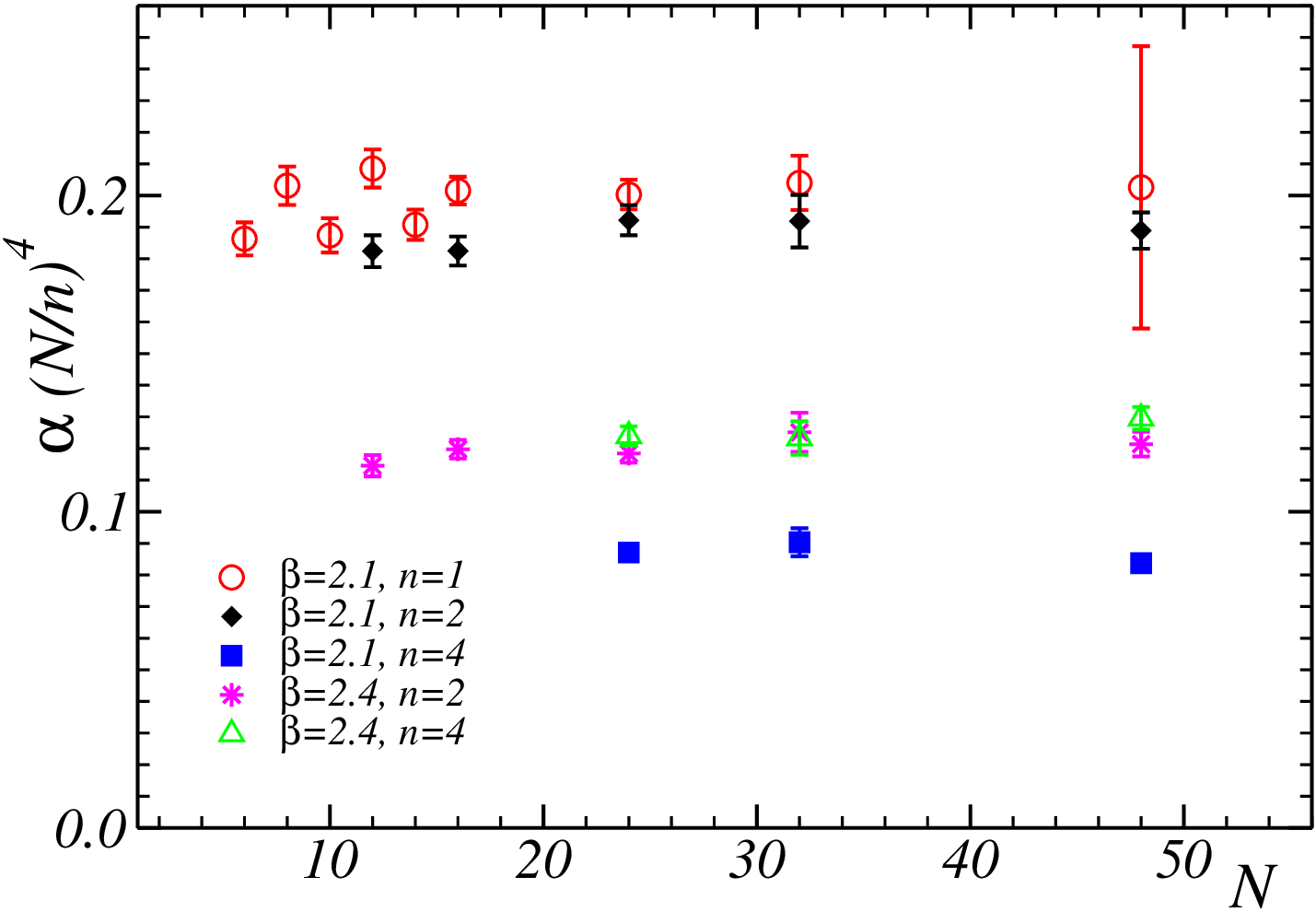} \\
(a) & (b)
\end{tabular}
\caption{(a) The fitting parameter $\alpha$ as the function of the size $N/n$ of the
coarse lattice; (b) the parameter $\alpha$ multiplied by the lattice volume as the
function of the lattice size $N$.}
\label{fig:alpha:vs:Nn}
\end{figure}

In order to confirm our expectation \eq{eq:alpha:b:V} we plot the parameter $\alpha$ $vs.$ the
ratio $N \slash n$ in Figure~\ref{fig:alpha:vs:Nn}(a) for selected set of coupling
constants $\beta$ and the blocking steps of the monopole, $n$. Since the volume of
the blocked lattice is ${(N \slash n)}^4$, we expect that the parameter $\alpha$ behaves as
$\alpha \propto {(N \slash n)}^{-4}$. This behaviour is seen in
Figures~\ref{fig:alpha:vs:Nn}(a,b). The parameter $\alpha$ multiplied by the lattice volume
almost does not depend on the lattice size $N$ according to Figure~\ref{fig:alpha:vs:Nn}(b).

\begin{figure}[thb]
\vskip 12mm
\includegraphics[angle=-0,scale=0.4]{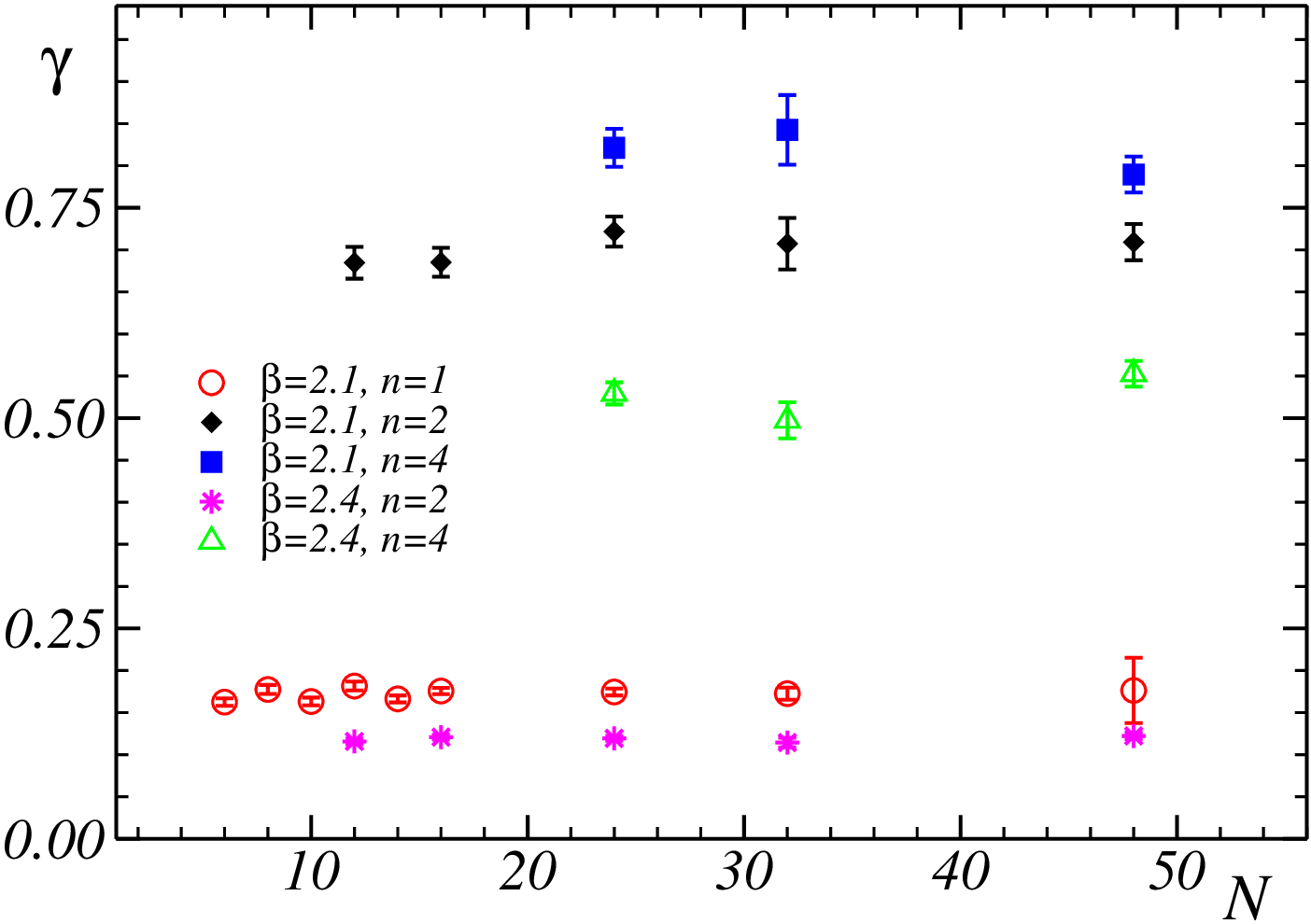}
\caption{The illustration of the independence of the fitting
parameter $\gamma$ on the lattice size, $N$.}
\label{fig:gamma:vs:N}
\end{figure}

According to our discussion above the fitting parameter $\gamma$ should only be a function
of the blocking size $b$ and should not depend on the volume of the lattice.
In Figure~\ref{fig:gamma:vs:N} we show the parameter $\gamma$ is indeed independent of the
lattice size $N$.

\begin{figure}[thb]
\vskip 12mm
\begin{tabular}{cc}
\includegraphics[angle=-0,scale=0.3]{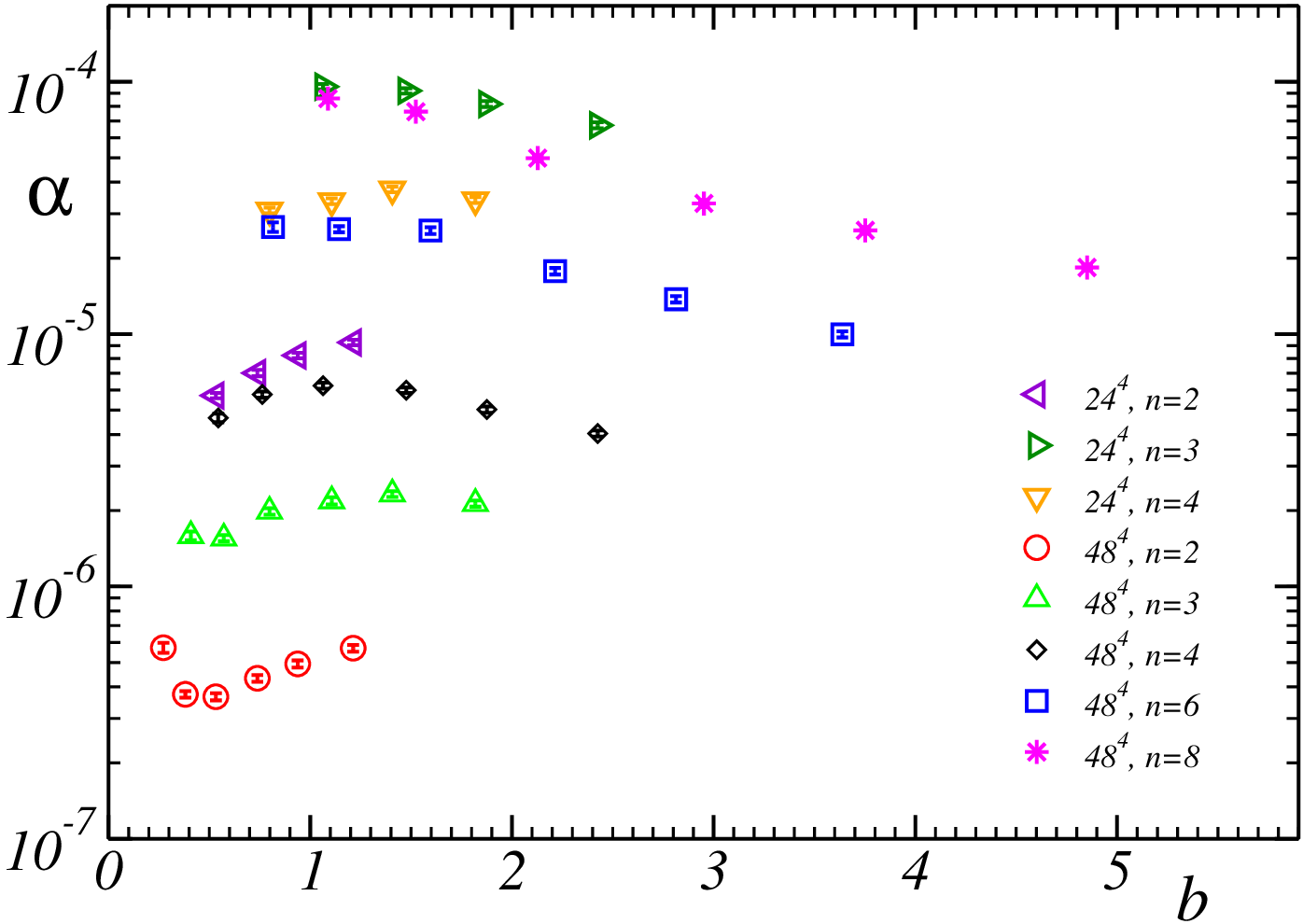} &
\hskip 12mm
\includegraphics[angle=-0,scale=0.3]{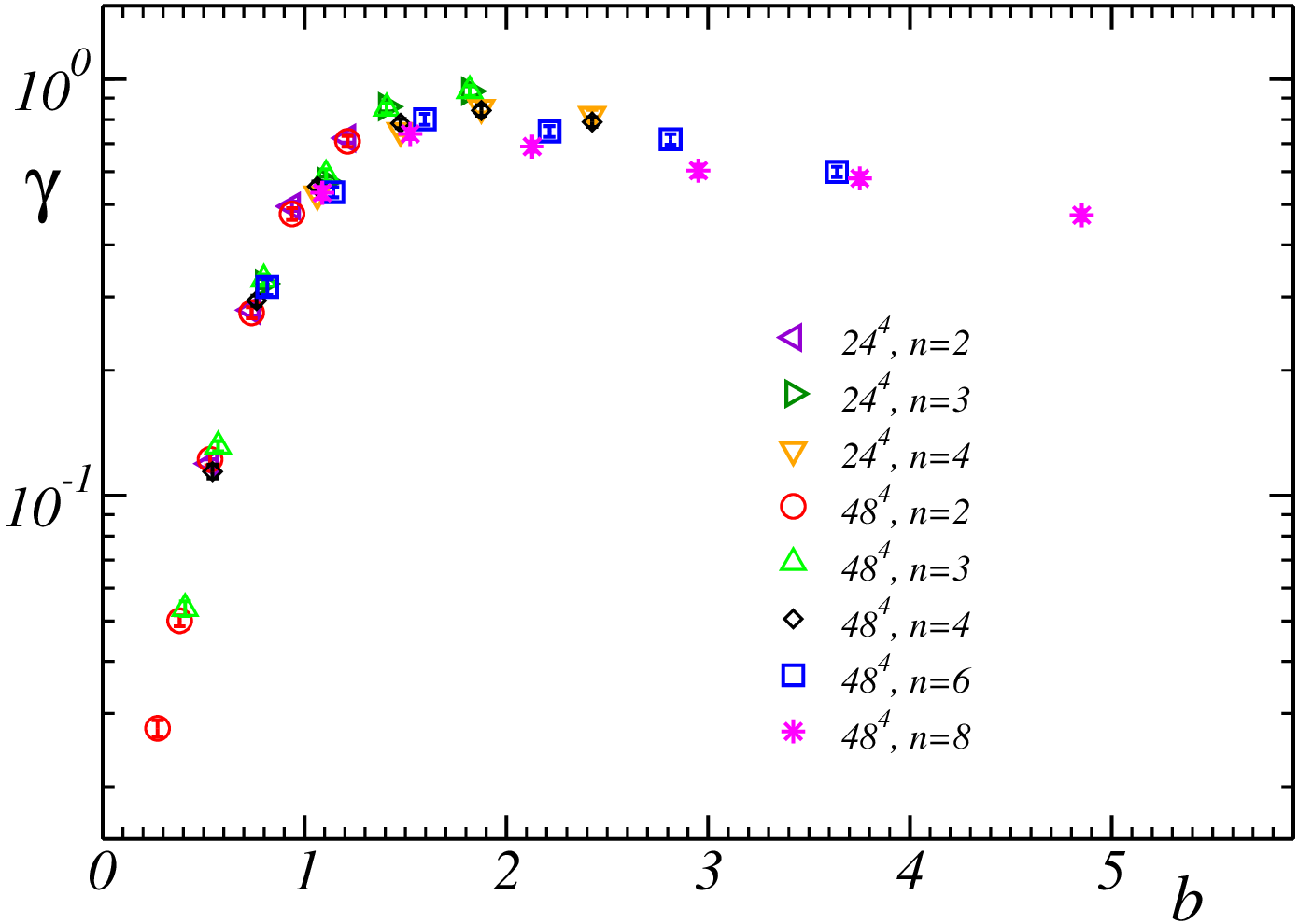}\\
(a) & (b) \\
\end{tabular}
\caption{The fitting parameters (a) $\alpha$ and (b) $\gamma$ as functions $b$ for
various lattice volumes, $N^4$, and monopole blocking steps, $n$.}
\label{fig:alpha:beta}
\end{figure}

The fitting parameters $\alpha$ and $\gamma$ are shown
as functions of the physical scale $b$ in Figures~\ref{fig:alpha:beta}(a) and (b),
respectively. The parameter $\gamma$ shows the scaling behaviour in a sense that
it depends on the blocking step $n$ and lattice spacing $a$ in the form of
the product $b=n\cdot a$.

\section{Monopole density and entropy}

\subsection{Monopole density}

The simplest physical characteristic of the monopole ensemble is its density.
It is interesting to compare the monopole density obtained from the IR monopole cluster
distribution, Eq.~\eq{eq:Lmax:2}, with the direct observation of the monopole density,
\beqn
\rho_{IR} = \frac{1}{4 {(n a)}^3 \cdot (N/n)^4} \Bigl \langle \sum_{s,\mu} |k^{(n)}_\mu(s)|
\Bigr\rangle\,.
\label{eq:mon:dens:Lmax}
\eeqn
Here the blocked monopole current, $k^{(n)}_\mu$, is defined by Eq.~\eq{eq:blocking}.
The normalization factor in Eq.~\eq{eq:mon:dens:Lmax} appears naturally if one
notes that $b = n a$ and $4 (N/n)^4$ are
the lattice spacing and the number of links of the coarse lattice, respectively.

\begin{figure}[thb]
\vskip 12mm
\begin{tabular}{cc}
\includegraphics[angle=-0,scale=0.3]{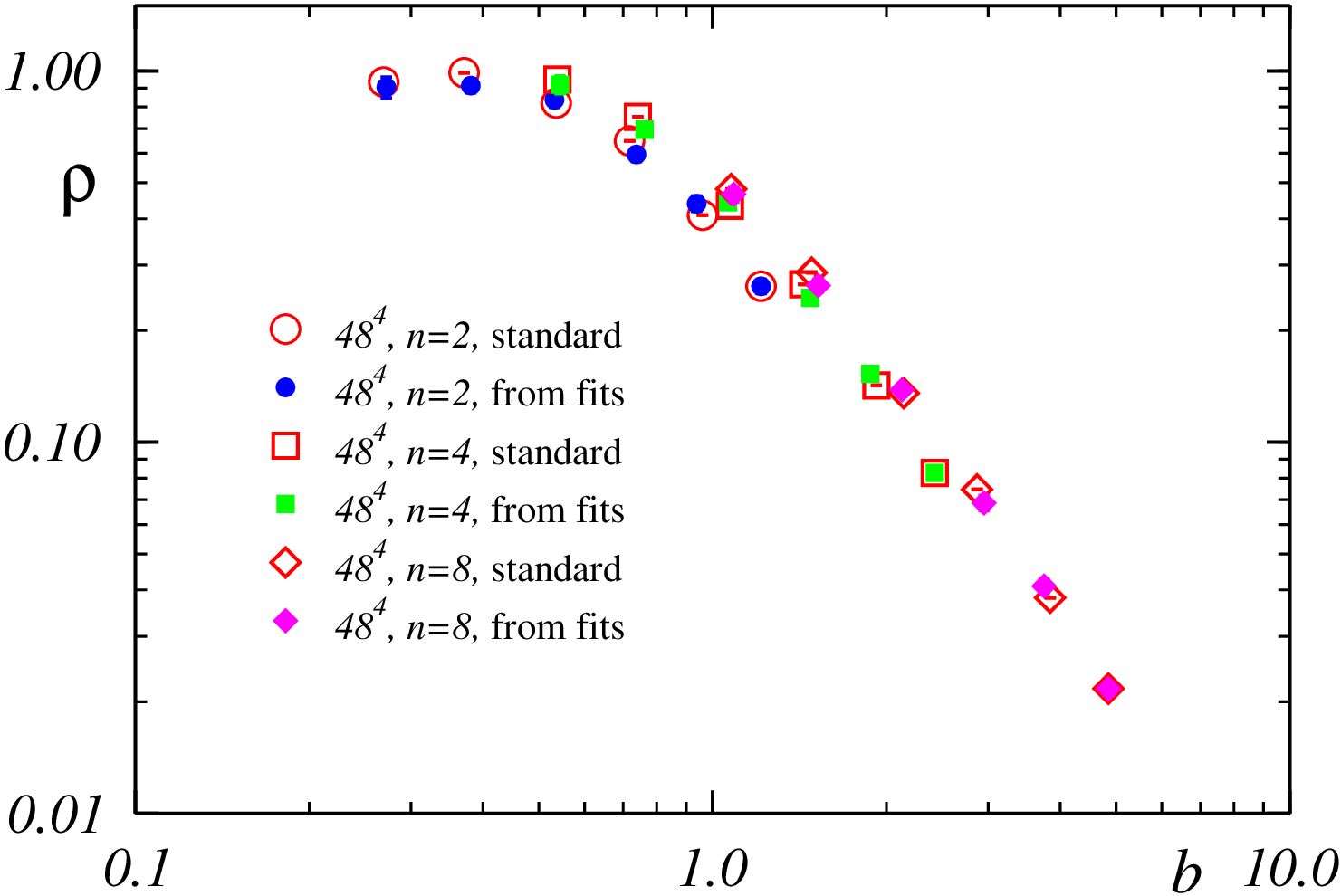} &
\hskip 12mm
\includegraphics[angle=-0,scale=0.3]{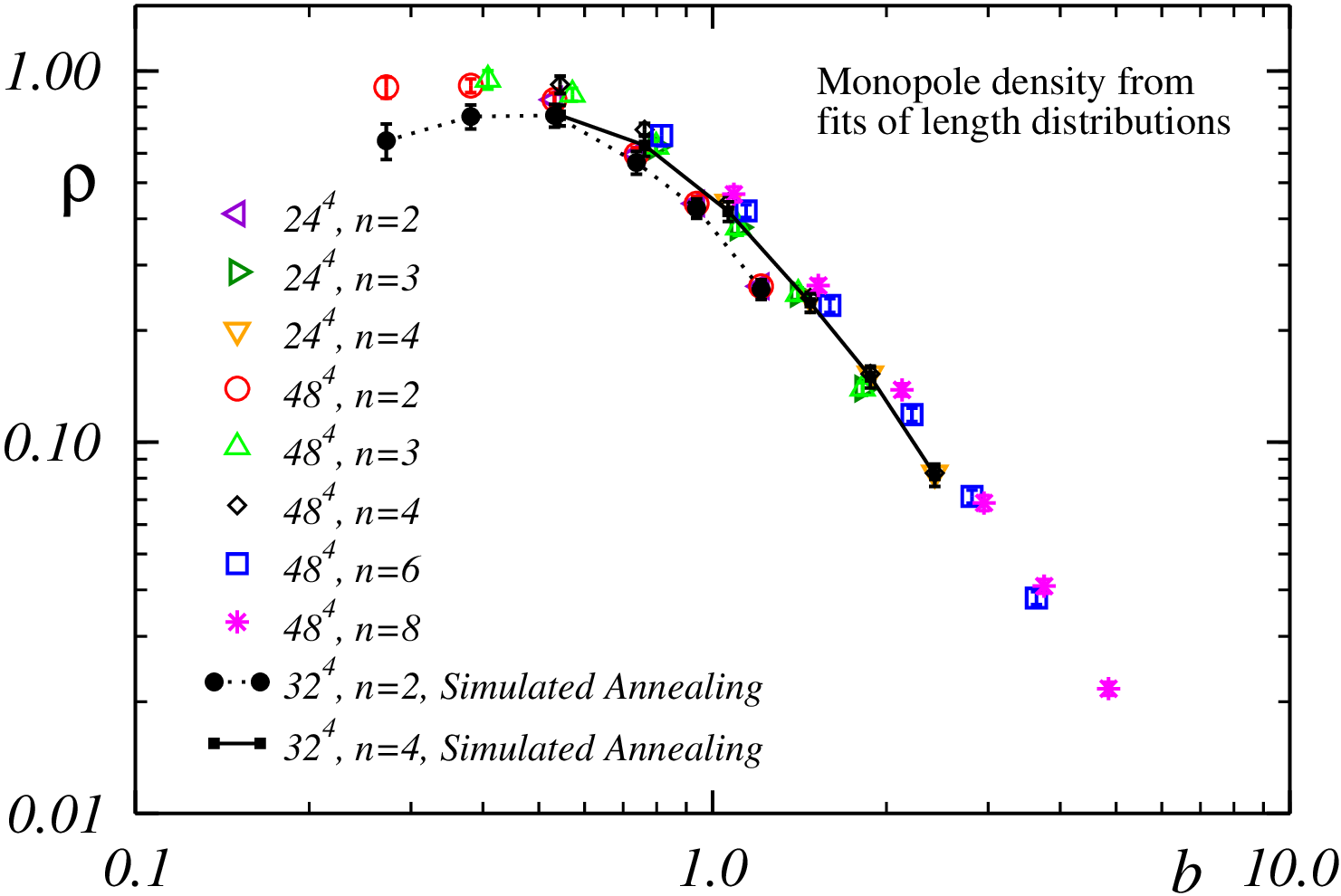}\\
(a) & (b) \\
\end{tabular}
\caption{(a) Comparison of the infrared monopole density obtained from
the fits of the monopole distributions, (\ref{eq:IR:distr:two},\ref{eq:Lmax:2})
with the density obtained in a direct
way~\eq{eq:mon:dens:Lmax}. (b) Comparison of the effect of the gauge
fixing procedure (iterative $vs.$ simulated annealing)
on the infrared monopole density.}
\label{fig:mon:dens}
\end{figure}

If the fitting function~\eq{eq:IR:distr:two} describes the data correctly then one
should observe no essential difference between the infrared monopole density
obtained from the fits of the monopole distributions, (\ref{eq:IR:distr:two},\ref{eq:Lmax:2})
compared to the density obtained in a direct way~\eq{eq:mon:dens:Lmax}. This is indeed the
case according to Figure~\ref{fig:mon:dens}(a).

Another information which can be extracted from this Figure is that
the blocked monopole density goes to a fixed limit, $\lim_{b \to 0}\rho
\approx 0.9\, \sigma^{3/2}$, as the blocking size $b$ gets smaller.
It is also possible that the monopole
density shows a wide plateau around $b \sqrt{\sigma} \approx 0.2$. In order to discriminate
between these options one should study the blocked monopole density at smaller values of
lattice spacing, $a$, and, consequently, at larger lattice volumes. We also note that the value of
the blocked monopole density quoted above is about $30\%$ larger than the value of
density~\cite{ref:bornyakov} of the {\it elementary} infrared monopoles in the continuum limit.

The monopole density is known to be sensitive to the details of the gauge fixing
procedure~\cite{ref:bornyakov}. In order to check the effect of the gauge fixing
we compare in Figure~\ref{fig:mon:dens}(b) the infrared monopole density obtained
using the SA and iterative gauge fixing algorithms. One can see from this Figure
that at large $b$ there is practically no difference between the monopole densities
obtained with the use of the different algorithms. However, there exists some difference
at small $b$ since the SA monopole density is smaller than the density obtained
with the help of the iterative algorithm. This slight dependence of the density
on the gauge fixing algorithm at small $b$ may explain the discrepancy between our results
and the results of Ref.~\cite{ref:bornyakov} mentioned above. Another source of the
discrepancy is the qualitative difference between the elementary and the blocked monopoles.
Since the scale $b$ is taken to be independent of the lattice spacing $a$ while $a$
tends to zero in the continuum limit, the elementary monopoles are expected to be more
affected by the ultraviolet lattice artifacts.

\subsection{Monopole entropy}

The distribution of the monopole trajectories depends both on the
monopole action and on the monopole entropy as we have already discussed in
Section~\ref{sec:length}. Therefore the knowledge of the distribution and
the monopole action allows us to define the entropy of the monopole currents.
If the monopoles make a simple random walk on the four--dimensional hypercubic lattice
then the entropy factor for elementary
monopoles is expected to be equal to seven, $\mu=7$, since there are seven choices
at each site for the monopole current to go further (the monopole trajectory is
obviously non--backtracking due to the presence of the magnetic charge).

The balance between energy and entropy of the elementary monopole trajectories
plays an important role. For example, the compact U(1) gauge model in four
dimensions possesses a phase transition associated with the monopole
(de)condensation which is defined as a point on the phase diagram where the
entropy and the energy of the monopole trajectories are the same. This point
is located by the condition $\gamma=0$, where $\gamma$
characterizes the monopole distribution, Eq.~\eq{eq:IR:distr:inf} or \eq{eq:IR:distr:two}.

In the case of the compact $U(1)$ gauge model the coefficient $f$ in the action of the monopole
trajectory, $S = f L$, is proportional to the lattice gauge coupling,
$f_{U(1)} \propto \beta_{U(1)}$ and $\mu=7$. This
fact allowed authors of Ref.~\cite{ref:BMK} to find the critical value of $\beta_{U(1)}$
deconfinement transition point with a great accuracy. Indeed, if $\gamma$ is negative then
the infrared cluster disappears and the confinement of electric charges is lost. The energy-entropy
balance was also studied numerically for the monopoles in compact U(1) gauge
theory~\cite{ref:suzuki:shiba} and in finite--temperature pure SU(2) gauge
theory~\cite{ref:kitahara}.

In the pure zero temperature QCD the coupling $\gamma$ is positive at all values of the lattice
coupling constant\footnote{The coupling $\beta$ must be smaller then a certain value at which
the unphysical deconfinement transition happens at finite volume.} $\beta$. The approximate
cancellation of the entropy factor and the energy of the elementary monopoles in
the zero--temperature gluodynamics
is known as "fine tuning"~\cite{ref:zakharov:clusters,ref:zakharov}. This fact
is in agreement with the existence of physical scaling of the infrared cluster in
zero--temperature pure QCD.

\begin{figure}[thb]
\vskip 12mm
\includegraphics[angle=-0,scale=0.4]{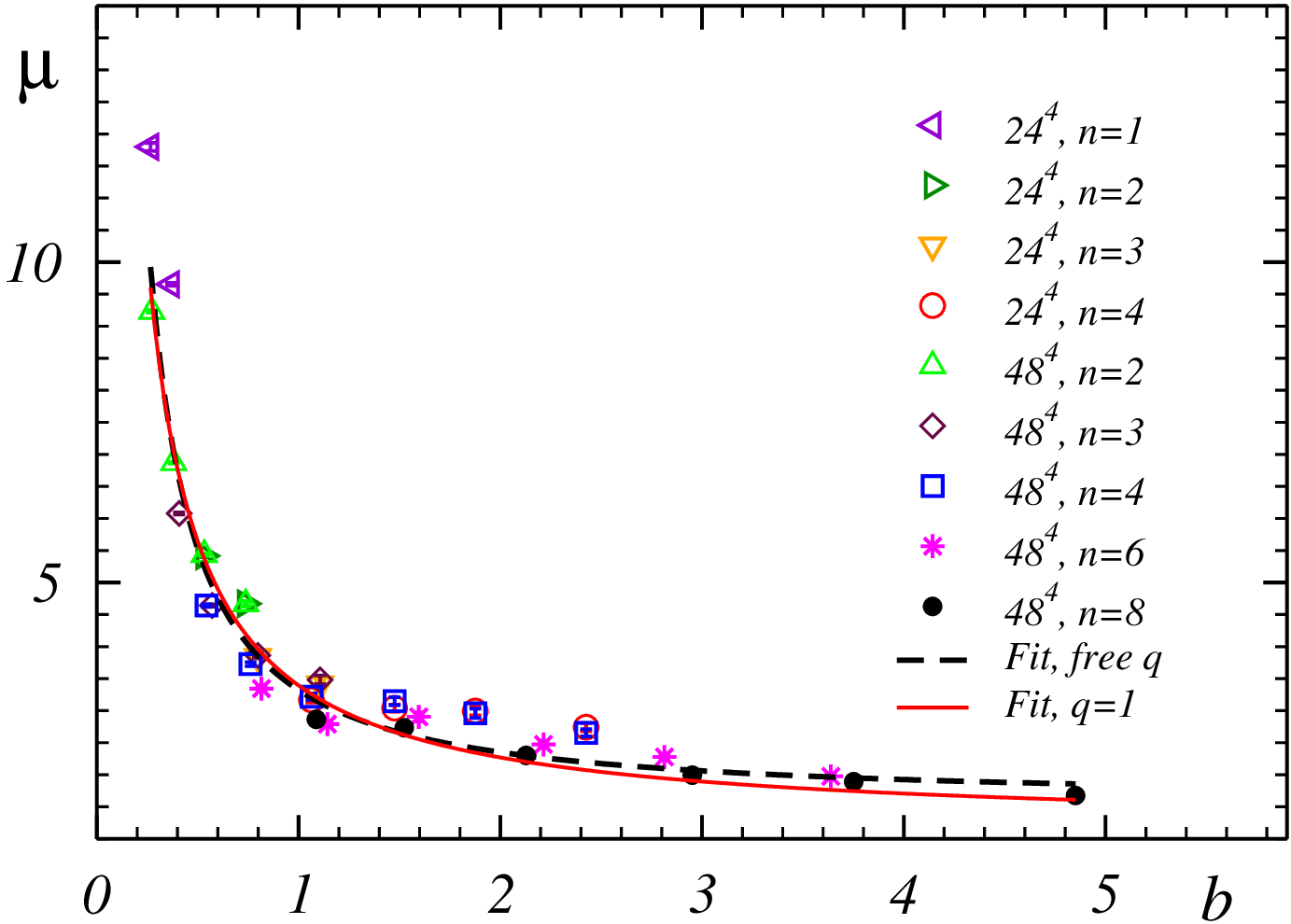}
\caption{Entropy factor $\mu$ $vs.$ $b$. The dashed line represents the fit by Eq.~\eq{eq:fit:entropy}
with free $q$--parameter and the solid line corresponds to the fixed parameter, $q=1$.}
\label{fig:entropy}
\end{figure}

The entropy factor $\mu$ of the infrared monopole trajectories can be
obtained from the IR cluster distribution and the monopole action according to
Eq.~\eq{eq:IR:distr:inf},
\beqn
\mu = e^{\gamma + f}\,.
\label{eq:mu}
\eeqn
We calculate numerically the parameters $\gamma$ and $f$ to find the entropy factor
$\mu$ for various scales $b$ and lattice sizes. Our results are presented in
Figure~\ref{fig:entropy}. The entropy shows an approximate scaling behaviour in a sense that
the entropy depends only on the scale $b$ and is independent of the lattice spacing, $a$, and the
blocking factor, $n$, separately. One can also notice that the entropy $\mu$ is
independent of the volume of the lattice. The largest scaling violations happens at small
blocking sizes $n=1,2$ at which the finite--size artifacts are expected to be strong.

The entropy factor $\mu$ is a declining function of the scale $b$. According to discussion above
one can expect that for elementary monopoles the factor $\mu$ should be equal to seven. However, $\mu>7$
for small values of $b$, as can be seen from Figure~\ref{fig:entropy}. We explain this small--$b$
behaviour as an artifact of our numerical procedure adopted in this paper. Indeed, we have used
the quadratic monopole action while at small $b$ higher--point interaction terms are essential and thus
the monopole action can not be reliably described by the quadratic terms only~\cite{chernodub}.

At large $b$ the entropy factor \eq{eq:mu} is smaller than seven. Formally this means that the
motion of the blocked monopoles is constrained. We have fitted the entropy by the function:
\beqn
\mu^{\mathrm{fit}} = \mu_{\infty} + C\, \mu^{-q}\,,
\label{eq:fit:entropy}
\eeqn
where $\mu_{\infty}$, $C$ and $q$ are the fitting parameters. The best fit is shown in
Figure~\ref{fig:entropy} by the dashed line. The corresponding best fit parameters are:
$\mu_{\infty} = 1.6(4)$, $C = 1.7(5)$ and $q = 1.2(2)$. The most interesting fitting
parameter is $\mu_{\infty}$ which is the asymptotic value of the entropy in the infrared
limit $b \sigma^{1/2}\to \infty$ according to Eq.~\eq{eq:fit:entropy}. Unfortunately, the
value of the asymptotic entropy is obtained with a big error bar in the above fit. In order to
increase the accuracy we notice that the power $q$ is very close to unity. Fixing $\mu=1$
in Eq.~\eq{eq:fit:entropy} and repeating the fitting procedure again we get
$\mu_{\infty} = 1.15(25)$ and $C = 2.2(1)$. The corresponding best fit curve is shown
in Figure~\ref{fig:entropy} by the solid line.

The fact that the asymptotic value of the entropy is very close to the unity in the
limit $b \sigma^{1/2}\to \infty$ may have a simple explanation. The monopole with a
large blocking size $b$ behaves as a classical object and its motion is no more a simple
random walk. The predominant motion of the large--$b$ monopole is close to a straight line.

\section{Conclusion}

We studied numerically the distributions of the infrared monopole currents of various blocking sizes,
$n$, on the lattices with different spacings, $a$, and volumes, $N$. The distributions can be described by
a gaussian anzatz with a good accuracy. The anzatz contains two important terms: (i) the linear term, which
possesses information about the energy and entropy of the monopole currents; and (ii) the quadratic term, which
suppresses too large infrared clusters. The linear term is independent of the lattice volume while the quadratic
term is inversely proportional to the volume. The monopole density determined from the parameters of
the gaussian fits coincides with the result of the direct numerical calculation.

We also studied the action of the monopoles belonging to the infrared clusters and compared it with the
action of the total monopole ensemble. It turns out that the self--coupling coefficients for both
these ensembles are almost the same at large $b$. However, as the blocking scale $b$ is decreased
the self--coupling coefficient for infrared monopole cluster gets noticeably larger then the
coefficient for the total monopole ensemble.
This can be explained by the fact that the self-interaction coefficient depends on the monopole density
(the larger density the smaller the coefficient is), and the difference between the total density and
the infrared density increases as $b$ gets smaller.

The knowledge of both the coefficient in front of the linear term of the gaussian distribution and the
monopole action for infrared clusters allows us to determine the entropy factor of the extended (blocked)
monopole currents. We have numerically shown that the entropy of the blocked monopole currents is a descending
function of $b=n a$, indicating that the effective degrees of freedom of the blocked monopoles are getting
smaller as the blocking scale $b$ increases. This corresponds to the classical picture: the monopole
with the large blocking size $b$ becomes a macroscopic object and the motion of such a monopole is
close to a straight line.

\begin{acknowledgments}
M.Ch. acknowledges the support by JSPS Fellowship No. P01023. T.S. is partially
supported by JSPS Grant-in-Aid for Scientific Research on Priority Areas
No.13135210 and (B) No.15340073.
\end{acknowledgments}

\end{document}